\documentclass[letter,11pt]{article}

\usepackage{setspace}
\usepackage{subcaption}
\usepackage{enumerate}
\usepackage{amsmath}
\usepackage{amsfonts}
\usepackage{graphicx}
\usepackage{tcolorbox}
\usepackage{amssymb}
\usepackage{multirow}
\usepackage{geometry}
\usepackage{soul}
\usepackage{colortbl}
\usepackage{wrapfig}
\usepackage{algorithm}
\usepackage{algorithmicx}
\usepackage{amsthm}
\usepackage{todonotes}
\usepackage{mathtools}
\usepackage{mathrsfs}

\usepackage[pdftex, plainpages = false, pdfpagelabels, 
                 bookmarks=false,
                 bookmarksopen = true,
                 bookmarksnumbered = true,
                 breaklinks = true,
                 linktocpage,
                 pagebackref,
                 colorlinks = true,  % was true
                 linkcolor = blue,
                 urlcolor  = blue,
                 citecolor = red,
                 anchorcolor = green,
                 hyperindex = true,
                 hyperfigures
                 ]{hyperref}

\usepackage{thmtools} 
\usepackage{thm-restate}

\newcommand{\capinstance}
{(U,\mathsf{cap},\mathcal{A})}

\newcommand{\cA}{\mathcal{A}}
\newcommand{\cF}{\mathcal{F}}
\newcommand{\cO}{\mathcal{O}}

\newcommand{\cP}{\mathcal{P}}
\newcommand{\capa}{\mathsf{cap}}
\newcommand{\score}{\mathsf{score}}

\newcommand{\cov}{\mathsf{cov}}

\newcommand{\cbra}[1]{\left\{ #1 \right\}}
\newcommand{\bin}{\{0,1\}}
\newcommand{\opt}{\mathsf{OPT}}

\newcommand{\val}{\mathsf{val}}

\algtext*{EndWhile}% Remove "end while" text
\algtext*{EndIf}% Remove "end if" text
\algtext*{EndFor}% Remove "end for" text
\algtext*{EndProcedure}
\algtext*{EndFunction}

\newtheorem{lemma}{Lemma}

\numberwithin{lemma}{section}
\newtheorem{theorem}[lemma]{Theorem}
\newtheorem{theorem*}[lemma]{Theorem*}

\newtheorem{definition}[lemma]{Definition}
\newtheorem{observation}[lemma]{Observation}

\newtheorem{claim}[lemma]{Claim}
\newtheorem{proposition}[lemma]{Proposition}

\usepackage{wrapfig}
\usepackage{bm}

\title{Parameterized Approximation for Capacitated $d$-Hitting Set with Hard Capacities}
\author{
	Daniel Lokshtanov\thanks{University of California Santa Barbara, USA.}
	\and
    Abhishek Sahu\thanks{National Institute of Science, Education and Research, An OCC of Homi Bhabha National Institute, Bhubaneswar, India.}
    \and
	Saket Saurabh \thanks{
Department of Informatics, University of Bergen, Norway}
\thanks{Institute of Mathematical Sciences, Chennai, India.}
\thanks{Saket Saurabh acknowledges support by European Research Council (ERC) under the European Union’s Horizon 2020 research
and innovation programme (grant agreement No. 819416), and Swarnajayanti Fellowship grant DST/SJF/MSA01/2017-18.
\begin{minipage}{0.1\textwidth}
    \begin{center}
        \includegraphics[scale=0.5]{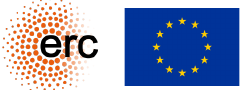}
    \end{center}
\end{minipage}} 
	\and
	Vaishali Surianarayanan\footnotemark[1]
    \and
    Jie Xue\thanks{New York University Shanghai, China.}
 }

\begin{document}

\pagenumbering{gobble}

\maketitle

\begin{abstract} 
In the \textsc{Capacitated $d$-Hitting Set} problem input is a universe $U$ equipped with a capacity function $\mathsf{cap}: U \rightarrow \mathbb{N}$, and a collection $\mathcal{A}$ of subsets of $U$, each of size at most $d$. The task is to find a minimum size subset $S$ of $U$ and an assignment $\phi :  \mathcal{A} \rightarrow S$ such that, for every set $A \in \mathcal{A}$ we have $\phi(A) \in A$ and for every $x \in U$ we have $|\phi^{-1}(x)| \leq \mathsf{cap}(x)$. Here $\phi^{-1}(x)$ is the collection of sets in $\mathcal{A}$ mapped to $x$  by $\phi$. Such a set $S$ is called a {\em capacitated hitting set}. When $d=2$ the problem is known under the name  \textsc{Capacitated Vertex Cover}. In  \textsc{Weighted Capacitated $d$-Hitting Set} each element of $U$ has a positive integer weight and the goal is to find a capacitated hitting set of minimum weight. 

Approximation algorithms for \textsc{Capacitated Vertex Cover} were first studied by Chuzhoy and Naor [SICOMP 2006], who gave a factor $3$ approximation algorithm for \textsc{Capacitated Vertex Cover} and showed that the weighted version does not admit an $o(\log n)$-approximation unless P=NP. 
%
%%
%%, and after a series of improvements 
%%\textsc{Capacitated Vertex Cover} and \textsc{Capacitated $d$-Hitting Set} are well understood from the perspective of polynomial time approximation algorithms.
%%
%%
After a series of improvements spanning a period of 15 years, Kao [SODA 2017] and Wong [SODA 2017] independently obtained $d$-approximation algorithms for \textsc{Capacitated $d$-Hitting Set}.
%%the best known algorithm for unweighted \textsc{Capacitated $d$-Hitting Set}, obtained simultaneously and independently by Kao [SODA 2017] and of Wong [SODA 2017], achieves an approximation ratio of $d$. %
%
This matches the ratio for the classic \textsc{$d$-Hitting Set} problem, and therefore cannot be improved to $d-\epsilon$ for any $\epsilon > 0$ assuming the Unique Games Conjecture. 
\textsc{Capacitated Vertex Cover} is also well understood from the perspective of parameterized algorithms: van Rooij and van Rooij [SOFSEM 2019] gave a $k^k|U|^{O(1)}$ time algorithm to determine whether there exists a solution $S$ of size at most $k$, showing that the unweighted problem is {\em fixed parameter tractable} (FPT) parameterized by the solution size $k$.

In this paper we initiate the study of parameterized (approximation) algorithms for \textsc{Capacitated $d$-Hitting Set}. An easy reduction shows that, as opposed to \textsc{Capacitated Vertex Cover}, unweighted \textsc{Capacitated $d$-Hitting Set} for $d \geq 3$ does {\em not} admit an FPT algorithm unless FPT=W[1]. 
Our main result is a parameterized approximation algorithm that runs in time 
$\left(\frac{k }{\epsilon} \right)^k 2^{k^{\cO(kd)}}(|U|+|\mathcal{A}|)^{O(1)}$ 
and either concludes that there is no solution of size at most $k$ or outputs a solution $S$ of size at most $4/3 \cdot k$ and weight at most $2+\epsilon$ times the minimum weight of a solution whose size is at most $k$.
We note that while the running time of our algorithm depends on $d$, the approximation ratio does not. 
Thus this parameterized approximation algorithm achieves a ratio smaller than what can be achieved by polynomial time approximation algorithms with a running time that is faster than what can be achieved by an algorithm solving the problem exactly. 
We complement our algorithmic results by showing that there exists a constant $c > 1$ such that for every $d \geq 3$, assuming the Exponential Time Hypothesis, there is no FPT-approximation algorithm for unweighted \textsc{Capacitated $d$-Hitting Set} with approximation factor $c$, and no factor $2-\epsilon$ FPT-approximation algorithm for weighted \textsc{Capacitated $d$-Hitting Set}. These hardness results even hold for \textsc{Capacitated Vertex   Cover} in multigraphs. On the way we show that a variant of multi-dimensional {\sc Knapsack} is hard to FPT-approximate within a factor $2-\epsilon$. This result may be of independent interest. 
\end{abstract}

\section{Introduction}
In the classic {\sc Vertex Cover} problem, the input is a graph $G$, and the task is to find a minimum size vertex subset $S$ such that every edge of $G$ has at least one endpoint in $S$. In the more general {\sc Hitting Set} problem input is a universe $U$, and a family ${\cal A}$ of subsets of $U$. The task is to find a minimum size subset $S$ of $U$ such that every set in ${\cal A}$ has non-empty intersection with $U$. When every set in ${\cal A}$ has size upper bounded by some fixed constant $d$, this problem is called $d$-{\sc Hitting Set}. {\sc Hitting Set} (and $d$-{\sc Hitting Set}) are also known under the name {\sc Hypergraph Vertex Cover} (in $d$-regular hypergraphs). Indeed, $2$-{\sc Hitting Set} is equivalent to {\sc Vertex Cover}.

Guha et al.~\cite{DBLP:journals/jal/GuhaHKO03}, motivated by an application in glycobiology, initiated the study of a capacitated version of $d$-{\sc Hitting Set}. 
In {\sc Capacitated $d$-Hitting Set} the input additionally contains a capacity function $\mathsf{cap} : U \rightarrow \mathbb{N}$ and a multiplicity function $M : U \rightarrow \mathbb{N}$.
The task is to find a subset $S$ of $U$
and an assignment $\phi : {\cal A} \rightarrow S$ such that 
$\textsf{size}(S,\phi) = \sum_{v \in S} \lceil \frac{|\phi^{-1}(v)|}{\mathsf{cap}(v)} \rceil$ is minimized
and for every $v \in U$ it holds that $\lceil \frac{|\phi^{-1}(v)|}{\mathsf{cap}(v)} \rceil \leq M(v)$.
Here $\phi^{-1}(v)$ is the subset of ${\cal A}$ that is mapped to $v$ by $\phi$.
Informally we can buy each element $v$ of $U$ up to $M(v)$ many times, and each time we buy $v$ we choose up to $\mathsf{cap}(v)$ sets in ${\cal A}$ that get covered by $v$. We now briefly consider the different variants of the problem that have been studied.

\smallskip 
\noindent {\bf Weighted vs Unweighted:} In {\sc Weighted Capacitated $d$-Hitting Set} input additionally comes with a weight function $w : (u) \rightarrow \mathbb{N}$ and the weight of a solution $(S, \phi)$ is $\textsf{weight}(S, \phi) = \sum_{v \in S} w(v) \cdot \lceil \frac{|\phi^{-1}(v)|}{\mathsf{cap}(v)} \rceil$. The task is to find a solution of minimum weight.

\smallskip
\noindent {\bf Soft vs Hard Constraints:} In the version of the problem with {\em soft constraints} $M(v) = \infty$ for every $v$, so each vertex can be bought an arbitrary number of times. The general case, with ``hard constraints'', is assumed by default unless stated otherwise. 

\smallskip
\noindent {\bf Simple Graphs vs Multigraphs:} By default we will allow ${\cal A}$ to be a {\em multi-family}, that is the same set $A$ can occur in ${\cal A}$ multiple times, and each occurrence of $A$ needs to be covered separately. In the {\em simple} version of {\sc Capacitated $d$-Hitting Set} each set in ${\cal A}$ is required to occur at most once, so ${\cal A}$ is a normal set family.

%the {\em weighted} version of the problem 
%Since several different variants of the problem have been studied, we define the

%By default we will allow ${\cal A}$ to be a {\em multi-family}, that is the same set $A$ can occur in ${\cal A}$ multiple times, and each occurrence of $A$ needs to be covered separately. In the {\em simple} version of {\sc Capacitated $d$-Hitting Set} each set in ${\cal A}$ is required to occur at most once, so ${\cal A}$ is a normal set family.
%

\smallskip
Guha et al.~\cite{DBLP:journals/jal/GuhaHKO03} obtain a factor $d$ approximation algorithm for {\sc Weighted Capacitated $d$-Hitting Set} with soft constraints. 
%
%Here $M(v) = \infty$ for every $v$, so each vertex can be bought an arbitrary number of times. 
%
Chuzhoy and Naor~\cite{DBLP:journals/siamcomp/ChuzhoyN06} initiated the study of {\sc Capacitated $2$-Hitting Set} (under the name {\sc Capacitated Vertex Cover}, and give a factor $3$ approximation for simple {\sc Capacitated Vertex Cover}.
They show that (simple) {\sc Weighted Capacitated Vertex Cover} is as hard to approximate as {\sc Hitting Set}, and therefore does not admit a $o(\log n)$-approximation unless $\mathsf{P}=\mathsf{NP}$~\cite{DBLP:journals/toc/Moshkovitz15}.
Subsequently Gandhi et al.~\cite{DBLP:journals/jcss/GandhiHKKS06} obtained a factor $2$ approximation algorithm for simple {\sc Capacitated Vertex Cover}.

Chuzhoy and Naor~\cite{DBLP:journals/siamcomp/ChuzhoyN06} posed as open problems whether it is possible to achieve a constant factor approximation for {\sc Capacitated $d$-Hitting Set} for every $d$, and whether the requirement that ${\cal A}$ be simple can be dropped.has
Both problems were resolved in the affirmative by Saha and Khuller~\cite{DBLP:conf/icalp/SahaK12}, who achieved a ratio of $38$ for {\sc Capacitated Vertex Cover} and $\max(6d, 65)$ for {\sc Capacitated $d$-Hitting Set} for $d \geq 3$, respectively. This naturally led to the question of whether it is possible to achieve a ratio for {\sc Capacitated $d$-Hitting Set} which comes close to (or matches) the (optimal, assuming the Unique Games Conjecture~\cite{khot2008vertex,DBLP:journals/siamcomp/DinurGKR05}) ratio of $d$ for {\sc $d$-Hitting Set}. Making progress towards this goal, Cheung, Goemans and Wong~\cite{DBLP:conf/soda/CheungGW14} gave algorithms with ratio $2.1555$ for $d=2$ and $2d$ for $d\geq 3$ respectively. 
The approximability of {\sc Capacitated $d$-Hitting Set} was settled completely by Kao~\cite{DBLP:conf/soda/Kao17,DBLP:journals/algorithmica/Kao21} and Wong~\cite{DBLP:conf/soda/Wong17}, who simultaneously and independently obtained $d$-approximation algorithms for the problem. 

The {\sc Capacitated Vertex Cover} problem in simple graphs with hard constraints $M(v) = 1$ for every vertex $v$ has also been studied from the perspective of parameterized algorithms~\cite{CyganFKLMPPS15,DBLP:series/txcs/DowneyF13}. Guo, Niedermeier and Wernicke~\cite{DBLP:journals/mst/GuoNW07}  gave an algorithm with running time $2^{O(k^2)}n^{O(1)}$ for finding a solution $S$ of size at most $k$, thereby showing that the problem is {\em fixed parameter tractable} (FPT) with respect to the parameter $k$. Dom et al.~\cite{DBLP:conf/iwpec/DomLSV08} gave an improved algorithm with running time $k^{3k}n^{O(1)}$, this was in turn further improved to $k^kn^{O(1)}$ by van Rooij and van Rooij~\cite{DBLP:conf/sofsem/RooijR19}. Prior to our work no versions of {\sc Capacitated $d$-Hitting Set} for $d \geq 3$ have been studied from the perspective of parameterized algorithms.
\\

We initiate the study of parameterized (approximation) algorithms for {\sc Capacitated $d$-Hitting Set} parameterized by the solution size $k$. An easy reduction from the {\sc Partial Vertex Cover} (the exact same reduction as Theorem 6.1 of~\cite{DBLP:conf/soda/CheungGW14}) shows that, unless \textsf{FPT} = \textsf{W[1]}, {\sc Capacitated $d$-Hitting Set} does not admit an FPT algorithm for every $d \geq 3$, even when there are no multi-edges, $M(x) = 1$ for every vertex $x$ and all weights are 1.
Our main result is a parameterized algorithm that simultaneously achieves an approximation ratio of $4/3$ in terms of the size of the solution, and an approximation ratio of $2+\epsilon$ for the weight. 
\\ \\
\begin{theorem}\label{thm:mainAlg}
There exists an algorithm that takes as input an instance
$(U, {\cal A}, \mathsf{cap}, M, w)$ of {\sc Weighted Capacitated $d$-Hitting Set}, an integer $k$ and rational $\epsilon > 0$,
runs in time $$\left(\frac{k }{\epsilon} \right)^k 2^{k^{\cO(kd)}}(n+m)^{\mathcal{O}(1)}$$ and either outputs that $(U, {\cal A}, \mathsf{cap}, M, w)$ has no solution of size at most $k$,
or outputs a solution $(S,\phi)$ with $\textsf{size}(S,\phi) \leq 4/3 \cdot k$ and $w(S,\phi) \leq (2 + \epsilon) \cdot w(S^\star, \phi^\star)$, where $(S^\star, \phi^\star)$ is a minimum weight solution to the instance out of all solutions of size at most $k$.
\end{theorem}
%\todo[inline]{abhishek vaishali please fill with correct running time when you have it}

The algorithm of Theorem~\ref{thm:mainAlg} works for the default version of {\sc Capacitated $d$-Hitting Set} with arbitrary vertex multiplicities and when ${\cal A}$ is allowed to be a multiset. The algorithm of Theorem~\ref{thm:mainAlg} is a $4/3$ approximation algorithm for unweighted instances, and a $(2+\epsilon)$ approximation for finding a solution whose weight is compared to the minimum weight instance of size at most $k$. It achieves both goals simultaneously with one solution. 
Note that the \textsf{W[1]}-hardness result for finding exact solutions of size at most $k$ in instances with $w(v) = 1$ for every  $v$ implies that any algorithm with running time $f(k,d)(|U||{\cal A}|)^{O(1)}$ must incur some loss {\em both} in the size and in the weight of the solution that it outputs.
The approximation guarantees achieved by our algorithm in FPT time are far better than what is possible for the problem if only polynomial time is allowed: $4/3$-approximation vs $(d-\epsilon)$ hardness for unweighted instances, and $2$-approximation vs $o(\log n)$-hardness for weighted ones. 

%The approximation ratio of $4/3$ of the algorithm of Theorem~\ref{thm:mainAlg} does not depend on $d$ and is better than what can be achieved even for {\sc Capacitated Vertex Cover}  (and even just {\sc Vertex Cover}) in polynomial time. Similarly the approximation ratio of $2$ for wei
%
%Similarly, the ratio of $2$ for weighted instances compares favorably
%\item 
%
%\item What do we say about weights? \todo{!!}

We complement our algorithm by showing that, assuming the Exponential Time Hypothesis~\cite{impagliazzo2001problems} (ETH), the $4/3$-approximation guarantee for unweighted instances cannot be made arbitrarily close to $1$, and that the ratio $2+\epsilon$ for weighted instances is essentially tight. We first state the FPT-inapproximability result for unweighted instances.

\begin{theorem}\label{main:hardnessUnweighted}
%There exists a constant $c > 1$\todo{explicit contant here?} such that for every $d \geq 2$, assuming the ETH there is no algorithm that takes as input an instance $(U, {\cal A}, \mathsf{cap}, M)$ of {\sc Capacitated $d$-Hitting Set} and integer $k$, runs in time $f(k,d)(|U||{\cal A}|)^{O(1)}$ and either outputs that there is no solution of size at most $k$ or outputs a solution of size at most $4/3 \cdot k$
There exists a constant $c > 1$
%ok \todo{explicit contant here?} 
such that for every integer $d \geq 2$, assuming the ETH there is no FPT (with parameter $k$) algorithm that distinguishes between instances $(U, {\cal A}, \mathsf{cap}, M, k)$ of {\sc Capacitated $d$-Hitting Set} that have a solution of size at most $k$ or have no solution of size at most $c \cdot k$.
\end{theorem}

The hardness result of Theorem~\ref{main:hardnessUnweighted} holds even when $M(v) = 1$ for every element $v$. For $d \geq 3$ it holds even for simple instances where $M(v) = 1$ for every element $v$.
Note that Theorem~\ref{main:hardnessUnweighted} implies that (assuming the ETH) there is no FPT algorithm for {\sc Capacitated Vertex Cover} in multigraphs, even when $M(v) = 1$ for every element $v$, in stark contrast to the FPT algorithms for unweighted simple graphs~\cite{DBLP:journals/mst/GuoNW07,DBLP:conf/iwpec/DomLSV08,DBLP:conf/sofsem/RooijR19}. Next, we state our FPT-inapproximability result for weighted instances.

\begin{theorem}\label{main:hardnessWeighted}
For every real $c < 2$ and integer $d \geq 2$, assuming the ETH, there is no FPT (with parameter $k$) algorithm that distinguishes between instances $(U, {\cal A}, \mathsf{cap}, M, w, k, W)$ of {\sc Weighted Capacitated $d$-Hitting Set} that have a solution of size at most $k$ and weight at most $W$ or have no solution of weight at most $c \cdot W$.
\end{theorem}
Similarly to Theorem~\ref{main:hardnessUnweighted}, the hardness result of Theorem~\ref{main:hardnessWeighted} holds even when $M(v) = 1$ for every element $v$. For $d \geq 3$ it holds even for simple instances where $M(v) = 1$ for every element $v$.

On the way to proving Theorems~\ref{main:hardnessUnweighted} and \ref{main:hardnessWeighted} we show hardness of FPT-approximating {\sc Multi-Dimensional Knapsack}~\cite{GURSKI201993}. Here input consists of $n$ vectors $v_1, \ldots, v_n \in \{0,1,\ldots,M\}^d$, a target vector $t \in \{0,1,\ldots,M\}^d$. The task is to select a smallest possible subset $S$ of $\{v_1, \ldots, v_n\}$ such that $\sum_{v_i \in S} v_i \geq t$. Thus this problem is just like the classic {\sc Knapsack} problem where (a) every item has unit cost, (b) we want to minimize the number of items to take such that their total value exceeds the given target value $t$, and (c) the target and values of the items are $d$-dimensional integer vectors instead of integers.  Towards proving Theorem~\ref{main:hardnessUnweighted} we show the following hardness result.

\begin{theorem}\label{thm:knapsackWeakHardness}
There exists a real $c > 1$ such that, assuming the ETH, there is no FPT (with parameters $k, d$) algorithm that distinguishes between instances $(v_1 \ldots v_n, t, k)$ of {\sc Multi-Dimensional Knapsack} with $d \leq 8.5k$ that have a solution of size at most $k$ or have no solution of size at most $c \cdot k$.
\end{theorem}

We need the constraint that $d = O(k)$ in the statement of Theorem~\ref{thm:knapsackWeakHardness} because our reduction (this reduction is very simple, and can be found in Section~\ref{sec:reductionToUnweightedCVC}) from {\sc Multi-Dimensional Knapsack} to {\sc Capacitated $d$-Hitting Set} introduces $d$ new elements that are forced into the solution to the {\sc Capacitated $d$-Hitting Set} instance. Hence a $c$-approximation for the constructed instances is allowed additive error terms that are linear in $d$.
When reducing to {\sc Weighted Capacitated $d$-Hitting Set} we can give these new elements weight $0$, and thus we no longer need the requirement that $d \leq O(k)$. Without this additional requirement we can prove a substantially stronger hardness result for  {\sc Multi-Dimensional Knapsack}.

\begin{theorem}\label{thm:knapsackStrongHardness}
For every real $c < 2$, assuming the ETH, there is no FPT (with parameters $k, d$) algorithm that distinguishes between instances $(v_1 \ldots v_n, t, k)$ of {\sc Multi-Dimensional Knapsack} that have a solution of size at most $k$ or have no solution of size at most $c \cdot k$.
\end{theorem}

The hardness results of Theorems~\ref{main:hardnessUnweighted},~\ref{main:hardnessWeighted},~\ref{thm:knapsackWeakHardness} and~\ref{thm:knapsackStrongHardness} are all proved by reduction from {\sc 3-Regular 2-CSP}.
Until very recently the parameterized inapproximability of {\sc 3-Regular 2-CSP} was only known under rather strong complexity assumptions~\cite{DBLP:conf/soda/LokshtanovR0Z20}. However, in a recent breakthrough result, Guruswami et al.~\cite{DBLP:conf/stoc/GuruswamiLRS024} showed hardness of approximating {\sc 3-Regular 2-CSP} assuming the ETH. 

\medskip

\noindent 
{\bf Related Work.} Over the past decade, a flurry of results has focused on the design of parameterized approximation algorithms for various combinatorial problems. 
Some of the notable problems that have been shown to admit FPT-approximation algorithms include  {\sc Vertex  Minimum Bisection}~\cite{FeigeM06}, {\sc$k$-Path Deletion}~\cite{Lee19}, {\sc Max $k$-Vertex Cover} in $d$-uniform hypergraphs~\cite{SkowronF17,Manurangsi19,DBLP:conf/soda/0001KPSS0U23,DBLP:conf/icalp/00020LS0U24}, {\sc $k$-Way Cut}~\cite{GuptaLL18,GuptaLL18a,KawarabayashiL20,LokshtanovSS20} and {\sc Steiner Tree} parameterized by the number of non-terminals~\cite{DvorakFKMTV18}.  

These are just a few representative examples, even the following list of recent work~\cite{ChitnisHK13,DemaineHK05,DvorakFKMTV18,BelmonteLM18,FeigeM06,GrandoniKW19,GuptaLL18,GuptaLL18a,KawarabayashiL20,Kortsarz16,Lampis14,Lee19,Manurangsi19,Marx04,Marx08,LokshtanovSS20,PilipczukLW17,SkowronF17,Wiese17}
is far from exhaustive. 
On the other hand, several basic problems have been shown to be hard even to FPT-approximate. The main ones include {\sc Set Cover}, {\sc Dominating Set}, {\sc Independent Set}, {\sc Clique}, {\sc Biclique} and {\sc Steiner Orientation}~\cite{ChalermsookCKLM17,SLM19, Wlodarczyk20,ChenL19,Lin18,Lin19,DBLP:journals/corr/abs-1909-01986,DBLP:conf/stoc/GuruswamiLRS024}. For a comprehensive overview of the state of the art on Parameterized Approximation, we refer to the recent  survey by Feldmann et al.~\cite{FeldmannSLM20}, and the surveys by Kortsarz~\cite{Kortsarz16} and by Marx~\cite{Marx08}.

\medskip
\noindent 
{\bf Organization of the paper.} In the next section (Section~\ref{sec:prelim}) we give some preliminary notations, and a lemma about dominating set in a bipartite graph. Section~\ref{sec:overview} gives an intuitive overview of our main algorithm (proof of Theorem~\ref{thm:mainAlg}). In Section~\ref{sec:mainalgo} we define everything formally and prove Theorem~\ref{thm:mainAlg}. Sections~\ref{sec:hardness-unwt} and ~\ref{sec:hardness-wt} give our lower bound results. In particular we prove Theorem~\ref{main:hardnessUnweighted} in Section~\ref{sec:hardness-unwt}, about unweighted version of \textsc{Capacitated $d$-Hitting Set}. This section also includes Theorem~\ref{thm:knapsackWeakHardness} about {\sc Multi-Dimensional Knapsack}. In
Section~\ref{sec:hardness-wt}, we give the lower bound for {\sc Weighted Capacitated $d$-Hitting Set} as well as a stronger result for {\sc Multi-Dimensional Knapsack}, proving Theorem~\ref{thm:knapsackStrongHardness} and Theorem~\ref{main:hardnessWeighted}. We conclude with some open questions and directions in Section~\ref{sec:conclusion}.

\section{Preliminaries}
\label{sec:prelim}
%\textcolor{red}{DEFINE $\odot$}
For a graph $G$, we use $V(G)$ and $E(G)$ to denote the set of vertices and the edges of $G$, respectively. 
%be a graph. 
For a vertex $v \in V(G)$, $N_G(v)$ denotes the set of neighbors of $v$ in $G$ and $N_G[v]$ = $N_G(v) \cup \{v\}$. For a set $X \subseteq V(G)$, $N_G(X) = \bigcup_{v \in X} N_G(v) \setminus X$, and $N_G[X] = N_G(X) \cup X$. When the graph $G$ is clear from context, we drop the subscript $G$, and simply write $N( \cdot)$ and $N[ \cdot]$. If $X \subseteq V(G)$, then we let $G[X]$ denote the subgraph induced by $X$. 
% If $Y \subseteq E(G)$, then we let $G[Y]$ denote the graph with the vertex set  $V(G)$ and the edge set $Y$. If $y \in E(G)$, then we use $Y \cap y$ as a shorthand for $Y \cap \{y\}$.
The {\em degree} of a vertex $v \in V(G)$ is the number of neighbors it has. This is denoted by ${\sf deg}(v)$. Let $[n]$ denote the set $\{1,\dots,n\}$.  

A formal definition of the unweighted version of the problem that we work with is provided below. The weighted and multiplicity case fairly easily reduced to the unweighted case.

\begin{tcolorbox}[colback=gray!5!white,colframe=gray!75!black]
        \textsc{Capacitated $d$-Hitting Set} \hfill \textbf{Parameter:} $k$
        \vspace{0.1cm} \\
        \textbf{Input:} A universe $U$ equipped with a capacity function $\mathsf{cap}: U \rightarrow \mathbb{N}$, and a collection $\mathcal{A}$ of subsets of $U$ each of which has size at most $d$. \\
        \textbf{Goal:} Find a subset $S \subseteq U$ of size at most $k$ and a map $\phi:\mathcal{A} \rightarrow S$ such that $\phi(A) \in A$ for all $A \in \mathcal{A}$ and $|\phi^{-1}(\{x\})| \leq \mathsf{cap}(x)$ for all $x \in S$.
\end{tcolorbox}

Let $\mathcal{A}'\subseteq \mathcal{A}$ and $a\in U$, we denote by $\mathcal{A}' \odot a$ the sets in $\mathcal{A}\in \mathcal{A}'$ that contain $a$, i.e. $\mathcal{A}'\odot a = \{A~:~A\in \mathcal{A'}, a \in A\}$.
%\mathcal{A}\cap a\neq \emptyset\}$.
%
Given a subset $S\subseteq U$, an assignment of $\mathcal{A}$ to $S$ is a mapping $\phi:\mathcal{A}\rightarrow S$ such that $|\phi^{-1}(x)|\leq \mathsf{cap}(x)$. Given an element $e\in S$ and a set $\cA'\subseteq \cA$, we define $\cov_{\phi}(e,\cA'):=|\phi^{-1}(e)\cap \cA'|$ as the number of sets in $\cA'$ that are mapped to $e$. We also extend this notation naturally to each set $S'\subseteq S$ as $\cov(S',\cA'):=|\phi^{-1}(S')\cap \cA'|$.

We prove a simple result on a special red-blue dominating set instance that we will use in our algorithm.

\begin{lemma}\label{domset}
Let $G=R_0\uplus B_0$ be a bipartite graph, where $|R_0|=r$, $|B_0|=b$, $r<2b$ and for all vertex $v\in B_0$, we have that  ${\sf deg}(v)\geq 2$. Then, there exists $D\subseteq R_0$ of size at most $(b+r)/3$, such that $N(D)=B_0$.
\end{lemma}
\begin{proof}
We prove the above by constructing a set $D$ with the desired properties. We include a vertex in each round into $D$ and modify the input instance as follows. 
If there is a vertex $r\in R_i$ with at least two neighbors in $B_i$, we include $r$ in $D$. Following this inclusion, we update the instance by deleting $r$ and all its neighbors from the instance, i.e., $R_{i+1}=R_i\setminus\{r\}$, $B_{i+1}=B_i\setminus N(r)$, $G_{i+1}=G[R_{i+1}\uplus B_{i+1}]$. Note that as long as $|R_i|\leq 2|B_i|$, there is always a vertex in $R_i$ with degree at least $2$. Suppose that the inclusion process ends after $a$ rounds. Then, $|R_{a+1}| \geq 2|B_{a+1}|$. Observe that the degree of a vertex in $B_0$ never changes throughout the process. 

Let $F_i$ be any two arbitrary vertices deleted in round $i$ from $B_0$ and $B'= ((\bigcup_{i=1}^{a} F_i)\cup B_{a+1})\subseteq B$. We denote the size of $B'$ by $b'$. Now, we extend $D$ to a  set of size $(b'+r)/3$ where $b'=|B'|$ and $N(D)=B_0$. For every vertex $v\in B_{a+1}$, we add one of its neighbor to $D$.
Note that $|B_{a+1}|=b'-2a$ (from construction) and $r-a=|R_{a+1}| \geq 2B_{a+1}$. Putting these two equations together we get, $r-a\geq 2(b'-2a)\implies a\geq (2b'-r)/3$.
But, $|D|\leq a + |B_{a+1}| \leq b'-a \leq (b'+r)/3$. This consequently bounds the size of $D$ by $(b+r)/3$. And, $N(D)=B_0$ as every vertex in $B_0\setminus B_{a+1}$ has at least one neighbor among all the vertices added to $D$ in the $a$ rounds, and every vertex in $B_{a+1}$ also has a neighbor in $D$ by construction.  
\end{proof}
%\section{A $\frac{4}{3}$-approximation algorithm}

\section{Technical Overview of $\frac{4}{3}$-approximation algorithm}
\label{sec:overview}
In this section we give a technical overview of our approximation algorithm. In the next section, we give a formal definition of the algorithm, argue its correctness and analyze the running time of the algorithm. 

Consider an unweighted instance $(U,\mathsf{cap},\mathcal{A})$ of \textsc{Capacitated $d$-Hitting Set}, without the multiplicity function.  Let $\opt$ denote a hypothetical solution to our problem, and $n=|U|$ and $m=|\mathcal{A}|$. Furthermore, let the covering map be  $\theta:\mathcal{A} \rightarrow \opt$ such that $\theta(A) \in A$ for all $A \in \mathcal{A}$ and $|\theta^{-1}(\{x\})| \leq \mathsf{cap}(x)$ for all $x \in \opt$. Our algorithm is recursive and, at any point in time, maintains a tuple that has entries describing the current state of the algorithm. The tuple contains the following elements.
\begin{enumerate}
\setlength{\itemsep}{-1pt}
    \item A partial solution $S \subseteq \opt$. That is, $S$ is consistent with the hypothetical solution $\opt$. 
    \item Equivalence classes of sets in $\cal A$ with respect to $S$. This is defined as follows. Given a set $S$, let $\mathcal{P}_d(S)$ denote the family of all subsets of $S$ of size at most $d$. For $E\in \mathcal{P}_d(S)$ let ${\cal A}_E$ be the sets $A \in \cA$ such that $A\cap S=E$. That is, ${\cal A}_E$ is an equivalence class.

The importance of equivalence classes comes from the following. For each set $A \in {\cal A}_E$ we have that $A\cap S=E$. This implies that each element in $E $ is present in all sets. So, if collectively $E \cap \opt$ are assigned $t$ sets of ${\cal A}_E$ by $\theta$, then it does not matter which element of $E \cap \opt$ takes care of each individual one of them. All that matters is that the elements in $E \cap \opt$ collectively have the capacity of $t$. This provides us with a lot of flexibility, as we can ignore which sets are assigned to individual elements of $E$ and remember only the number of sets in ${\cal A}_E$ assigned to each element of $E$.

  % \todo[inline]{Talk about equivalence class and its flexibility since each has same neighborhood}
    
    \item A function $\pi:\cP_d(S)\rightarrow S$ that records which element of $S$ {\em covers} the most sets in the equivalence class ${\cal A}_E$. Let us call this {\em plurality function}. %\todo{DL: in future versions, rename to plurality function}
    
    \item Given the function $\pi$, we automatically get a partition of the equivalence classes based on which element of $S$ covers the most sets in the equivalence class ${\cal A}_E$.  Observe that the set of equivalence classes that are mapped by $\pi$ to the same element $s$ of $S$ can be viewed as a star whose central vertex is $s$ and the leaves are the equivalence classes mapped to $s$ by $\pi$ (so $s$ covers the greatest number of sets in the equivalence class out of all elements in $S$).
    % See Figure~\ref{figure:stars}.

%     \begin{figure}[t]
% \includegraphics[scale=.75]{diagram-20240617.pdf}
% \caption{Stars with centers in $S$ and equivalence classes as leaves.} \label{figure:stars}\todo{please use symbols accxorindg to the paper}
% \end{figure}
    Formally, this is stated as follows. For each $s\in S$, $$\cA^{\pi}_s:=\{A\in \cA_E: \pi(E)=s, E\in \cP_d(S)\}.$$ 
Let us remark that if $s$ is a central vertex and ${\cal A}_E\in \cA^{\pi}_s$ is an equivalence class corresponding to a leaf, then this does {\em not} mean that some other vertex not in $S$ could not cover more. It just means that among the elements of $S$, $s$ covers the most.  Formally, all this is captured in Definition~\ref{definition:A_E-A_s} in Section~\ref{sec:mainalgo}.
\end{enumerate}

So, our tuple currently consists of $(S,\pi)$. Now we will introduce a partition of $U\setminus S$ into our tuple. Toward this, we employ the {\em color coding technique} of Alon, Yuster, and Zwick~\cite{alon1994color}. That is, uniformly at random we give each element of $U\setminus S$ a color from $[r]$, where $r=k-|S|$. We denote the elements colored with the color $i$ by $X_i$. In a \emph{good} coloring, each color class $X_i$ contains exactly one element from $\opt \setminus S$. The probability of a coloring being \emph{good} is at least $\frac{1}{e^r}$ (see~\cite{CyganFKLMPPS15} for details). We directly obtain a deterministic algorithm by constructing a family of about $e^r$ different coloring functions, and iterating over all of them. This introduces another entry into our tuple. That is, $(S,X=X_1\uplus \cdots \uplus X_{k-|S|},\pi)$. Given the partition $X_1\uplus \cdots \uplus X_{k-|S|}$, we try to {\em guess} how an element $x\in (X_i\cap \opt)$ affects the current equivalence classes. In our allowed time, we can neither guess the exact sets $x$ covers nor guess how many sets in each of the equivalence classes $x$ covers. In fact, the number of such guesses can be as bad as $n^{2^{|S|}}$. To reduce this number, we apply the bucketing trick. 

For each element $x\in (X_i\cap \opt)$, $i \in [r]$, and $x\in S$,  we estimate its approximate coverage in each of the equivalence classes. That is, for each $x\in (X_i\cap \opt)$, $i \in [r]$, and $x\in S$, we estimate $|\theta^{-1}(\{x\})|$. 
Let the buckets be 
$$ \left\{\left\lceil \left(1+\frac{1}{3k}\right)^p \right\rceil ~:~ p \in \{1,\ldots,t=\lceil\log_{1+1/3k}m\rceil\}\right\}$$
Specifically, for each equivalence class ${\cal A}_E$, $E\in \mathcal{P}_d(S)$, we guess a value $p$ where $0 \leq p \leq t$, such that $x$ covers {\em at least}  $\left\lceil \left(1+\frac{1}{3k}\right\rceil\right)^p $ many sets and at most  $\left\lceil \left(1+\frac{1}{3k}\right\rceil\right)^{p+1}$ sets. Our estimate of the approximate coverage allows for an error up to a factor of $\left(1+\frac{1}{3k}\right)$. Our estimates are encapsulated in  the following function. 

$$\gamma: ([k-|S|] \cup S) \times \mathcal{P}_d(S) \rightarrow \{\lceil (1+\frac{1}{3k})^p \rceil: p \in \{1,\ldots,\lceil\log_{1+1/3k}|\cA|\rceil\}\}$$

We extend the definition of $\gamma$, slightly abusing the notation,  to the domain $([k-|S|]\cup S) \times S$ by defining for each $i\in [k-|S|]\cup S$ and $s\in S$, $$\gamma(i,s):=\sum_{E\in \pi^{-1}(s)}\gamma(i,E).$$
That is, $\gamma(i,s)$ estimates the approximate total coverage of an element $x\in (X_i\cap \opt)$ to the equivalence classes in $\cA^{\pi}_s$. In simple language,  $\gamma(i,s)$ estimates the approximate total coverage by an element $x\in (X_i\cap \opt)$ of the leaves of the star for which $s$ is a central vertex.  This brings us the notion of an annotated tuple $(S,X=X_1\uplus \cdots \uplus X_{k-|S|},\pi,\gamma)$. All this is formalized in the next definition.

\begin{definition}\label{definition:annotated-tuple-overview}
    An annotated tuple is a tuple $(S,X=X_1\uplus \cdots \uplus X_{k-|S|},\pi,\gamma)$ where
    \begin{enumerate}
        \item[(1)] $S,X\subseteq U$ with $S\cup X=U$, $|S|\leq k$ and $X\cap S=\emptyset$
        \item[(2)] $\pi: \mathcal{P}_d(S)\rightarrow S$
        \item[(3)] $\gamma: ([k-|S|] \cup S) \times \mathcal{P}_d(S) \rightarrow \{\lceil (1+\frac{1}{3k})^p \rceil: p \in \{1,\ldots,\lceil\log_{1+1/3k}|\cA|\rceil\}\}$
    \end{enumerate}
    We extend $\gamma$ to the domain $([k-|S|]\cup S) \times S$ by defining for each $i\in [k-|S|]\cup S$ and $s\in S$, $$\gamma(i,s):=\sum_{E\in \pi^{-1}(s)}\gamma(i,E)$$
    %E\in \cP_d(S):\pi(E)=s
    %
    % For each $E\in \cP_d(S)$ and $s\in S$, we define 
    % $$\cA_E:=\{A\in \cA: A\cap S=E\}$$
    % $$\cA_s:=\{A\in \cA_E: \pi(E)=s, E\in \cP_d(S)\}$$
    %
    Given $S$ and $X$ satisfying (1), we denote by $\cF_{S,X}$ the set of all possible annotated tuples $(S,X,\pi,\gamma)$

    %
    %When $S=\emptyset$, $\cP_d(S)=\emptyset$ and so $\pi$ and $\gamma$ have an empty domain or co-domain. We denote the annotated tuple in this case by $(\emptyset,X,.,.)$
\end{definition}

% The number of annotated tuples $(S,X,\pi,\gamma)$ is upper bounded by $$ 1 \times e^k \times |S|^{2^{|S|}} \times p^{k\times 2^{|S|}}. $$
% We first observe that we do not need to consider all the sets in $\mathcal{P}_d(S)$ when defining the equivalence classes. Indeed, each set in $\cA$ has size at most $d$ and hence its intersection with $S$ is upper bounded by $d$. Thus the only sets in $\mathcal{P}_d(S)$ which are relevant has size at most $d$. This immediately brings the size of the relevant sets in $\mathcal{P}_d(S)$ to $\sum_{j=0}^d {|S| \choose d }\leq |S|^d$. Thus, the number of annotated tuples $(S,X,\pi,\gamma)$ can be upper bounded by 

The number of annotated tuples $(S,X,\pi,\gamma)$ is upper bounded by $$ 1 \times e^k \times |S|^{2^{|S|}} \times p^{k\times 2^{|S|}}. $$
Observe that each set in $\cA$ has size at most $d$ and hence its intersection with $S$ is upper bounded by $d$. Thus we defined $\mathcal{P}_d(S)$ to be the family of all subsets $S$ of size at most $d$ and used it to define the equivalence classes. 
This immediately shows the size of $\mathcal{P}_d(S)$ to be $\sum_{j=0}^d {|S| \choose d }\leq |S|^d$. Thus, the number of annotated tuples $(S,X,\pi,\gamma)$ can be upper bounded by 
\begin{equation}
\label{equation:annotate-tuple-bound}
\begin{aligned}
     1 \times e^k \times k^{k^d} \times p^{k^{d+1}} \leq e^k \times k^{k^d} \times \lceil\log_{1+1/3k}m\rceil^{k^{d+1}} \leq k^{\cO(k^{d+1})}+m
\end{aligned}
\end{equation}

% \begin{definition} \label{definition:A_E-A_s}
%     Given a set $S\subseteq U$ and a function $\pi:\cP_d(S)\rightarrow S$, we define:
%     \begin{itemize}
%         \item For each $E\in \cP_d(S)$, $\cA_E:=\{A\in \cA: A\cap S=E\}$
%         \item For each $s\in S$, $\cA^{\pi}_s:=\{A\in \cA_E: \pi(E)=s, E\in \cP_d(S)\}$
%         \todo[inline]{explain in words what this function and set is doing -- connect it to what you explain in the words earlier}
%     \end{itemize}
% \end{definition}
Having defined the annotated tuple, we look for a solution that respects our choices. This can be formulated in the next definition. Recall that,  given an element $e\in S$ and a set $\cA'\subseteq \cA$, we define $\cov_{\phi}(e,\cA'):=|\phi^{-1}(e)\cap \cA'|$ as the number of sets in $\cA'$ that are assigned to $e$. We also extend this notation naturally to each set $S'\subseteq S$ as $\cov(S',\cA'):=|\phi^{-1}(S')\cap \cA'|$.

\begin{definition}
\label{definition:good-tuple-intro}
    An annotated tuple $(S,X,\pi,\gamma)$ is good for the hypothetical solution $\opt$ and the corresponding covering function $\theta$ if 
    \begin{enumerate}
    \setlength{\itemsep}{-1pt}
        \item $S\subseteq \opt$
        %\item $S\cap A\neq \emptyset$ for each $A\in \cA$
        %\item For each $E\in \cP_d(S)$, $\cov_{\phi^*}(S,E)\geq (1-\frac{1}{f(k)})|E|$
        %$|\phi^{-1}(S)\cap E|\geq (1-\frac{1}{f(k)})|E|$
        \item For each $i\in [k-|S|]$, $|X_i\cap \opt|=1$
         \item For each $i\in [k-|S|]$ and $E\in \mathcal{P}_d(S)$, $\gamma(i,E)=\lceil(1+1/3k)^p\rceil$, $p=\lfloor \log_{1+\frac{1}{3k}} \cov_{\theta}(\opt \cap X_i,\cA_E) \rfloor$
        \item For each $s\in S$ and $E\in \mathcal{P}_d(S)$, $\gamma(s,E)=\lceil(1+1/3k)^p\rceil$, $p=\lfloor \log_{1+\frac{1}{3k}} \cov_{\theta}(s,\cA_E) \rfloor$
        \item For each $E\in \mathcal{P}_d(S)$, $\pi(E)=\max_{s\in S}\cov_{\theta}(s,\cA_E)$
        % \item For each $i\in [k-|S|]$ and $E\in \mathcal{P}_d(S)$, $\gamma(i,E)=\lfloor \log_{1+\frac{1}{k}} |\phi^{-1}(S^*\cap X_i)\cap \cA_E| \rfloor$
        % \item For each $s\in S$ and $E\in \mathcal{P}_d(S)$, $\gamma(s,E)=\lfloor \log_{1+\frac{1}{k}} |\phi^{-1}(s)\cap \cA_E| \rfloor$
        % \item For each $E\in \mathcal{P}_d(S)$, $\pi(E)=\max_{s\in S}|\phi^{-1}(s)\cap \cA_E|$
    \end{enumerate}
    We also say $(S,X,\pi,\gamma)$ respects the solution $\opt$ and assignment $\theta$.
\end{definition}

% \begin{definition}\todo{if not needed remove}
% \label{definition:good-tuple-intro}
%     An annotated tuple $(S,X,\pi,\gamma)$ is good if there exists a solution $S^*$ to $\capinstance$ along with an assignment $\phi^*:\mathcal{A}\rightarrow S^*$ such that
%     \begin{enumerate}
%         \item $S\subseteq S^*$
%         %\item $S\cap A\neq \emptyset$ for each $A\in \cA$
%         %\item For each $E\in \cP_d(S)$, $\cov_{\phi^*}(S,E)\geq (1-\frac{1}{f(k)})|E|$
%         %$|\phi^{-1}(S)\cap E|\geq (1-\frac{1}{f(k)})|E|$
%         \item For each $i\in [k-|S|]$, $|X_i\cap S^*|=1$
%          \item For each $i\in [k-|S|]$ and $E\in \mathcal{P}_d(S)$, $\gamma(i,E)=\lceil(1+1/3k)^p\rceil$, $p=\lfloor \log_{1+\frac{1}{3k}} \cov_{\phi^*}(S^*\cap X_i,\cA_E) \rfloor$
%         \item For each $s\in S$ and $E\in \mathcal{P}_d(S)$, $\gamma(s,E)=\lceil(1+1/3k)^p\rceil$, $p=\lfloor \log_{1+\frac{1}{3k}} \cov_{\phi^*}(s,\cA_E) \rfloor$
%         \item For each $E\in \mathcal{P}_d(S)$, $\pi(E)=\max_{s\in S}\cov_{\phi^*}(s,\cA_E)$
%         % \item For each $i\in [k-|S|]$ and $E\in \mathcal{P}_d(S)$, $\gamma(i,E)=\lfloor \log_{1+\frac{1}{k}} |\phi^{-1}(S^*\cap X_i)\cap \cA_E| \rfloor$
%         % \item For each $s\in S$ and $E\in \mathcal{P}_d(S)$, $\gamma(s,E)=\lfloor \log_{1+\frac{1}{k}} |\phi^{-1}(s)\cap \cA_E| \rfloor$
%         % \item For each $E\in \mathcal{P}_d(S)$, $\pi(E)=\max_{s\in S}|\phi^{-1}(s)\cap \cA_E|$
%     \end{enumerate}
%     We also say $(S,X,\pi,\gamma)$ respects the solution $S^*$ and assignment $\phi^*$.
% \end{definition}

So, our objective is given a good annotated tuple $(S,X,\pi,\gamma)$ find a solution.  Towards this, we first employ a filtering step. 

\smallskip
\noindent 
{\bf Filtering  the candidate elements in each $X_i$.} 
In this step we go through each element $x \in X_i$, $i\in [r]$, and see whether the $\capa(x)$ satisfies the desired coverage requirement given by function $\gamma$. That is, for each $i\in [k-|S|]$, $$X_i' := \{x \in X_i: \mathsf{cap}(x) \geq \sum_{s \in S} \gamma(i,s) \text{ and } |\mathcal{A}_E \odot x| \geq \gamma(i,E) , \text{ for each } E \in \cP_d(S)\}$$

 An element in $X_i'$ is known as a \emph{candidate} element.  We will further refine the sets $X_i'$ based on other covering criteria. Let $X'=\cup_{i\in[r]} X_i'$.  The next filtering step will consider an element $v\in X_i'$ and a star centered on $s\in S$. It is based on the question: {\em how many additional sets $v$ can cover among the sets $\cA^{\pi}_s$? }  Towards this we will define a score $\score(v,s)$ for the vertex. For each $E\in \cP_d(S)$, a vertex $v$ can cover at most $|\cA_E \odot v|$ sets, that is, the number of sets in $\cA_E$, which $v$ is part of. Furthermore, the function $\gamma$ provides an upper bound on the number of sets $v$ should cover in $\cA_E$. Thus, for $v\in X'$ with $v\in X'_i$,

    $$\mathsf{n}(v,E):=
    \min\bigg(
    \bigg\lceil 1+\frac{1}{3k} \bigg\rceil \cdot \gamma(i,E),
    |\cA_E \odot v|\bigg) \text{ for each $E\in \cP_d(S)$}$$
    
    $$\mathsf{n}(v,s):=\sum_{E\in \pi^{-1}(s)} \mathsf{n}(v,E) \text{ for each $s\in S$}$$

To obtain a lower bound on the number of additional sets $v$ could cover among the sets $\cA^{\pi}_s$, we remove the lower bound on the number of sets it must cover among other stars from its capacity. This gives us $\score(v,s)$
    
    $$\score(v,s):=\min\bigg(\mathsf{n}(v,s),\capa(v)-\sum_{s'\in S, s'\neq s} \gamma(i,s)\bigg) \text{ for each $s\in S$}$$
Intuitively, $\score(v,s)$ denotes how many additional sets in $\cA^{\pi}_s$, $v$ could cover, if we select $v\in X_i'$ in our solution. 

For each star $s\in S$, we guess two numbers from $[r]$, the color of the vertices in $\opt$ that makes the highest and second highest contribution, to the star centered on $s$. In particular, we do as follows. Let $x_i$ be an element of $X_i' \cap \opt$. Furthermore, for all $i\in[r]$, let $\beta_i=|\theta^{-1}(x_i)\cap \cA^{\pi}_s |$. Sort the numbers $\beta_i$ in decreasing order and choose the indexes of the first two numbers. We capture (guess) these by two functions $\tau_1,\tau_2:S\rightarrow [k-|S|]$. This leads to the definition of an extended annotated tuple.  

\begin{definition}\label{annotatedTuple:overview}
    An extended annotated tuple is a tuple $(S,X,\pi,\gamma,\tau_1,\tau_2)$ such that:
    \begin{enumerate}
    \setlength{\itemsep}{-1pt}
        \item $(S,X,\pi,\gamma)$ is an annotated tuple
        \item $\tau_1,\tau_2:S\rightarrow [k-|S|]$
    \end{enumerate}
\end{definition}

We are ready to define the candidate sets in each of the color class $X_i'$, $i \in [r]$. For each $s\in S$, we sort elements in the  class $X_i'$ based on the $\score(v,s)$. Let $T(i,s)$ denote the first $dk^{10}$ elements in the sorted order. If $|X_i'|\leq k\cdot dk^{10}$, then $X_i''=X_i'$, else $X''_i:=\bigcup_{s\in \tau_{1}^{-1}(i)}T(i,s)$. Observe that in the second case we {\em do not include}  top elements in $X''$ corresponding to all stars. We only add elements corresponding to those stars for which the highest contribution comes from the element in the class. This is crucial and will be useful in our algorithm. This is encapuslated formally in the next definition. 

\begin{definition}\label{definition:candidate-set-overview}
    Given an extended annotated tuple $(S,X,\pi,\gamma,\tau_1,\tau_2)$, we define the candidate set of $(S,X,\pi,\gamma,\tau_1,\tau_2)$ to be the subset $X'':=X''_1\uplus \cdots \uplus X''_{k-|S|}$ of $X$ such that:
    \begin{itemize}
        \item  For each $i\in [k-|S|]$, let $T(i,s)$ be the top $dk^{10}$ vertices of $X'_{i}$ when sorted in non-increasing order of $\score(v,s)$. If $|X'_i|\leq dk^{11}$ let $X''_i:=X'_i$ and if not $X''_i:=\bigcup_{s\in \tau_{1}^{-1}(i)}T(i,s)$.
    \end{itemize}
\end{definition}

At this stage, we branch on the fact that $\opt$ intersects one of the elements in $X_i''$, $i\in [r]$, or $\opt$ is disjoint from all the selected elements. In the first case, we make progress as the size of $S$ increases and we can restart. In the disjoint case (which is more interesting), we greedily find a solution of the desired size. Observe that each element of $X_i''$, $i\in [r]$ has a higher score than the corresponding element of $\opt$ in the set $X_i''$. The only reason $\opt$ does not pick the elements we have kept is that the sets they may see have a large overlap. To compensate for the overlap, we select additional elements that lead to the size of the solution $\frac{4}{3}k$.

We first show how we can select two additional vertices in each of the color classes $X_i''$ such that it leads to a factor $2$  solution. Toward this, define a conflict graph $\cal C$ as follows. The vertex set $V({\cal C})=\cup_{i\in [r]}X_i''$. We add an edge between two vertices $u,v\in V({\cal C})$ if there is some vertex $s\in S$ such that 
$$|\cA^{\pi}_s  \odot u \cap |\cA^{\pi}_s \odot v| \geq \frac{1}{\rho=k^{4}}\min\bigg(|\cA^{\pi}_s  \odot u |, |\cA^{\pi}_s \odot v|\bigg).$$
That is, there is a significant overlap of these two vertices in a star. Taking advantage of the fact that every color class is substantially large as a function of $k$ and $d$, and recognizing that the graph $H$ exhibits sparsity, being $(kd)^{\cO(1)}$ degenerate (where each subgraph of $\cal C$ contains a vertex with a degree no greater than $(kd)^{\cO(1)}$), we demonstrate that it is possible to choose an independent set $I$ in $\cal C$ such that exactly two vertices from each color class $X_i''$, $i \in [r]$, are included. One can show that $I \cup S$ is a solution of size at most $2k$.

We improve on this bound as follows. We find an independent set, as above, but we do not select all the vertices in our solution.  
\begin{itemize}
 \setlength{\itemsep}{-1pt}
    \item If there exists a color class $i$, such that $|\tau^{-1}_1(i)|\geq 2$, then we select both elements in $I$ from this class to $S$, and discard the stars corresponding to 
   $\tau^{-1}_1(i)$.  
 \item So assume that for every class $i$,  $|\tau^{-1}_1(i)|\leq 1$.   In this case, we construct a bipartite graph $H:=(S\uplus [k-s],E)$ where $E:=E_1\cup E_2$, $E_1:=\{(s,\tau_1(s)):s\in S\}$ and $E_2:=\{(s,\tau_2(s)):s\in S\}$. That is, $V(H)$ is partitioned into $S$ and a set that has a vertex for each of the color classes. For each vertex $s\in S$ we add edges to the classes that contributes highest and second highest to the star centered on $s$. In this case, we find a set $D\subseteq [r]=[k-s]$ such that $N(D)=S$ of size $\frac{k}{3}$ using Lemma~\ref{domset}.
\end{itemize}
%\todo[inline]{check which all stars are discarded in the first step}

We can combine both steps and show that the resulting solution has size at most $\frac{4}{3}k$. Now we give an intuitive reason for the correctness of the algorithm. Fix an equivalence class $\mathcal{A}_E$, $E\in \cP_d(S)$. 
\begin{itemize}
\setlength{\itemsep}{-1pt}
    \item We lose a factor of $1+\frac{1}{3k}$ when we guessed $\gamma$ approximately. 
    \item Secondly, we lose because the elements that we select overlap a lot. 
\end{itemize}
The way we handle losses is as follows. For each equivalence class, we have two elements that try to take care of this loss. Since our selected elements are independent in ${\cal I}$, it implies that their intersection is limited and their union is huge. It is in our control to limit the intersection and increase the union by appropriate choices of $\gamma$ and $\rho$ in the definition of conflict graph. In fact, we achieve the desired property by selecting different orders of magnitude of losses and gains as a function of $k$ in $\gamma$ and $\rho$.  The running time of the algorithm can be shown to be $f(k,d)\cdot (n+m)^{\cO(1)}$.

\section{A $\frac{4}{3}$-approximation Algorithm}\label{sec:mainalgo}

% \textbf{Technical Overview:}
% \begin{itemize} 
%     \item Talk about equivalence class and its flexibility since each has same neighborhood
%     \item 90\% coverage by $S$
%     \item Two from each color class with independence notion
%     \item Talk about stars and their use to improve approximation factor
%     \item Now talk about top and second top to obtain similar guarantees for notS in every star
% \end{itemize}
%First we reduce \textsc{Capacitated $d$-Hitting Set} to a new auxiliary problem that we define below.
%

Our algorithm is recursive and, at any point, has a partially constructed solution $S$. We make some guesses to include an additional vertex to $S$ and solve the problem recursively. We first define a notion that captures the form of the guesses we make and thus the subproblems we solve. Given a set $S$, let $\mathcal{P}_d(S)$ denote the family of all subsets of $S$ of size at most $d$. 
Suppose $S$ is a partial solution, then we store information with respect to each $E\in \cP_d(S)$. Let $A_E$ be the sets $A \in \cA$ such that $A\cap S=E$. We store (i) an estimate of how much each vertex in the solution covers in $A_E$ and (ii) the element of $S$ that contributes the maximum to $A_E$. We also store a partition of $U\setminus S$ such that each part contains at most one solution vertex. We formalize the required notions and the subproblem we solve below. 

%\todo{figure for annotated tuple and good annotated tuple}
\begin{definition} \label{definition:A_E-A_s}
    Given a set $S\subseteq U$ and a function $\pi:\cP_d(S)\rightarrow S$, we define:
    \begin{itemize}
        \item For each $E\in \cP_d(S)$, $\cA_E:=\{A\in \cA: A\cap S=E\}$
        \item For each $s\in S$, $\cA^{\pi}_s:=\{A\in \cA_E: \pi(E)=s, E\in \cP_d(S)\}$
    \end{itemize}
\end{definition}
When $\pi$ is clear from context, we use $\cA_s$ instead of $\cA^{\pi}_s$.
\begin{definition}\label{definition:annotated-tuple}
    An annotated tuple is a tuple $(S,X=X_1\uplus \cdots \uplus X_{k-|S|},\pi,\gamma)$ where
    \begin{enumerate}
        \item[(1)] $S,X\subseteq U$ with $|S|\leq k$ and $X\cap S=\emptyset$
        \item[(2)] $\pi: \mathcal{P}_d(S)\rightarrow S$
        \item[(3)] $\gamma: ([k-|S|] \cup S) \times \mathcal{P}_d(S) \rightarrow \{\lceil (1+\frac{1}{3k})^p \rceil: p \in \{1,\ldots,\lceil\log_{1+1/3k}|\cA|\rceil\}\}$
    \end{enumerate}
    We extend $\gamma$ to the domain $([k-|S|]\cup S) \times S$ by defining for each $i\in [k-|S|]\cup S$ and $s\in S$, $$\gamma(i,s):=\sum_{E\in \pi^{-1}(s)}\gamma(i,E)$$
    %E\in \cP_d(S):\pi(E)=s
    %
    % For each $E\in \cP_d(S)$ and $s\in S$, we define 
    % $$\cA_E:=\{A\in \cA: A\cap S=E\}$$
    % $$\cA_s:=\{A\in \cA_E: \pi(E)=s, E\in \cP_d(S)\}$$
    %
    Given $S$ and $X$ satisfying (1), we denote by $\cF_{S,X}$ the set of all possible annotated tuples $(S,X,\pi,\gamma)$

    %
    %When $S=\emptyset$, $\cP_d(S)=\emptyset$ and so $\pi$ and $\gamma$ have an empty domain or co-domain. We denote the annotated tuple in this case by $(\emptyset,X,.,.)$
\end{definition}

\begin{definition}
\label{definition:good-tuple}
    An annotated tuple $(S,X,\pi,\gamma)$ is good if there exists a solution $S^*$ to $\capinstance$ along with an assignment $\phi^*:\mathcal{A}\rightarrow S^*$ such that
    \begin{enumerate}
        \item $S\subseteq S^*$
        %\item $S\cap A\neq \emptyset$ for each $A\in \cA$
        %\item For each $E\in \cP_d(S)$, $\cov_{\phi^*}(S,E)\geq (1-\frac{1}{f(k)})|E|$
        %$|\phi^{-1}(S)\cap E|\geq (1-\frac{1}{f(k)})|E|$
        \item For each $i\in [k-|S|]$, $|X_i\cap S^*|=1$
         \item For each $i\in [k-|S|]$ and $E\in \mathcal{P}_d(S)$, $\gamma(i,E)=\lceil(1+1/3k)^p\rceil$, $p=\lfloor \log_{1+\frac{1}{3k}} \cov_{\phi^*}(S^*\cap X_i,\cA_E) \rfloor$
        \item For each $s\in S$ and $E\in \mathcal{P}_d(S)$, $\gamma(s,E)=\lceil(1+1/3k)^p\rceil$, $p=\lfloor \log_{1+\frac{1}{3k}} \cov_{\phi^*}(s,\cA_E) \rfloor$
        \item For each $E\in \mathcal{P}_d(S)$, $\pi(E)=\max_{s\in S}\cov_{\phi^*}(s,\cA_E)$
        % \item For each $i\in [k-|S|]$ and $E\in \mathcal{P}_d(S)$, $\gamma(i,E)=\lfloor \log_{1+\frac{1}{k}} |\phi^{-1}(S^*\cap X_i)\cap \cA_E| \rfloor$
        % \item For each $s\in S$ and $E\in \mathcal{P}_d(S)$, $\gamma(s,E)=\lfloor \log_{1+\frac{1}{k}} |\phi^{-1}(s)\cap \cA_E| \rfloor$
        % \item For each $E\in \mathcal{P}_d(S)$, $\pi(E)=\max_{s\in S}|\phi^{-1}(s)\cap \cA_E|$
    \end{enumerate}
    We also say $(S,X,\pi,\gamma)$ respects the solution $S^*$ and assignment $\phi^*$.
\end{definition}

\begin{observation}\label{observation:exists-good-tuple}
Given a solution $S^*$, an assignment $\phi^*$, sets $S\subseteq S^*$ and $X\subseteq U$ with $X=X_1\uplus \cdots \uplus X_{k-|S|}$, $X\cap S=\emptyset$, $|S|\leq k$, and $|X_i\cap S^*|=1$ for each $i\in [k-|S|]$, there exists an annotated tuple $(S,X,\pi,\gamma)\in \mathcal{F}_{S,X}$ that respects $S^*$ and $\phi^*$.
%
%$(\emptyset,X,.,.)$ is good if there exists a solution $S^*$ with $|X_i\cap S^*|=1$ for each $i\in [k]$.
\end{observation}

Our main crux is encapsulated in the next lemma. 
\begin{lemma}
\label{lemma:annotated-d-hs}
    There exists an algorithm that takes as input an instance $\capinstance$, an annotated tuple $(S,X,\pi,\gamma)$ and in time $2^{k^{\cO(d)}}(n+m)^{\mathcal{O}(1)}$ either returns $\mathsf{FAIL}$, or succeeds and returns a solution $R$ to $\capinstance$ of size at most $\frac{4}{3}k$ such that $|R\cap X_i|\leq 2$ for each $i\in [k-|S|]$.
    Furthermore if $(S,X,\pi,\gamma)$ respects a solution $S^*$ of size $k$, the algorithm succeeds.
\end{lemma}

% \begin{theorem}
% \label{theorem:main-alg}
%     There exists an algorithm to \capdhs\ that takes as input an instance $\capinstance$ and in time $f(k)poly(n)$ either returns a solution of size at most $\frac{4}{3}k$ or determines there exists no solution of size at most $k$.
% \end{theorem}
We are now ready to prove our main Theorem using Lemma~\ref{lemma:annotated-d-hs}.
\begin{proof}[\textbf{Proof of Theorem~\ref{thm:mainAlg}}]
The algorithm does the following:
\begin{enumerate}
    \item We first reduce the instance $(U,\capa,\cA,M,w)$ to a weighted instance $(U',\capa',\cA',w')$:
        \begin{itemize}
            \item For each $x\in U$, let $c_v:=\min(k,M(v))$. We add $c_v$ many copies $x_1,\cdots,x_{c_v}$ of $x$ to $U'$ with each having $\capa'(x_i):=\capa(x)$ and weight $w'(x_i):=w'(x)$. Then for each set $A\in \cA$ such that $x\cap A\neq \emptyset$, we add $A'=\bigcup_{x\in A} \cup_{i\in [c_v]}x_i$. 
        \end{itemize}
    \item We use a $\log n$-approximation from~\cite{DBLP:journals/siamcomp/ChuzhoyN06} to obtain an estimate $W$ on the weight of the optimum solution of size at most $k$.
    \item Construct an $(n,k)$-perfect hash family $\mathcal{F'}$ of size ${e^k k^{\mathcal{O}(\log k)}} \log n$ using the algorithm from~\cite{CyganFKLMPPS15}. Further use $\mathcal{F'}$ to construct a corresponding hash family $\mathcal{F}$ for $U$ by ordering the elements in $U$ and mapping the integers to elements.
    \item For each possible function $\eta:[k]\rightarrow [\frac{k\log |U'|}{\epsilon}]$ and family $\{X_1,\cdots,X_k\}\in \mathcal{F}$
    \begin{itemize}
        \item For each $X_i$, $i\in [k]$ define $X'_i=\{x\in X_i:\eta(i)\leq w'(x)\leq \eta(i)+\frac{\epsilon W}{k\log |U'|}\}$   
        \item Let $X'=X'_1\uplus\cdots \uplus X_k$. For each $(\emptyset,X',\pi,\gamma)\in \mathcal{F}_{\emptyset,X'}$ invoke Lemma~\ref{lemma:annotated-d-hs} with $(\emptyset,X',\pi,\gamma)$ and $(U',\capa',\cA')$. If the Lemma returns a set $S'$ for any input, then construct the multiset $S=\{x:x_i\in S'\}$ and return $S$.
    \end{itemize}
    \item Return $\mathsf{FAIL}$ if no set was returned in the previous step.
    % \item Solve the instance approximately and obtain a solution $S^*$ of weight $w$.
    % \item Guess weight buckets and filter.
    % \item For each $\{X_1,\cdots,X_k\}\in \mathcal{F}$ use Lemma~\ref{lemma:annotated-d-hs} with $(\emptyset,X,.,.)$. If the algorithm returns a solution $S'$ of size at most $\frac{4}{3}k$, then return $S'$.  
\end{enumerate}

For ease of presentation, we assume that we are seeking a solution of size exactly $k$. For at most $k$, we can run the algorithm for each $\ell\leq k$. We now show that if the instance admits a solution of size $k$, our algorithm returns one with size at most $\frac{4}{3}k$ and weight $2\epsilon W^*$.

For the correctness, first observe that if there exists a solution $R$ to $(U,\capa,\cA,M,w)$ then it can be converted to a solution $R'$ to $(U',\capa',\cA',w')$ by taking for each element $x$ in $U$ occurring in $R$, the copies corresponding to its multiplicity in $R$ in $R'$. This can be done in reverse too, given $R'$ we can construct $R$ - note that this is what we do in Step 4.

Let $R^*=\{v^*_1,\cdots,v^*_k\}$ be an optimum solution to $(U',\capa',\cA',w')$ of weight $W^*$. Then the estimated weight $W\leq \log n W^*$. 
We define $\eta^*:[k]\rightarrow [\frac{k\log |U|}{\epsilon}]$ as $\eta^*(i):=w'(v^*_i)$ for each $i\leq k$. Further let $\{X_1,\cdots X_k\}$ be a set of sets in $\mathcal{F}$ with $|X_i\cap R^*|=1$. Since $\mathcal{F}$ was constructed from a $(n,k)$ hash family $\mathcal{F}'$, such a set exists.

%Assume there exists a solution $S^*$ to $\capinstance$ of size at most $k$. Then by construction of $\mathcal{F}$, we know that there exists a $\{X_1,\cdots,X_k\}\in \mathcal{F}$ such that $X_i\cap S^*=v^*_i$ for each $i\in k$. 

Let $X'$s be the ones constructed when the algorithm goes over $\eta^*$ and $\mathcal{F}$.
Observe that due to the bound on $W$ and the definition of $\eta^*$ and $X'$, $X'_i\cap S^*=v^*_i$. That is the solution is not filtered out by $X'$.

Further, let $\phi^*$ be an assignment corresponding to $R^*$. By Observation~\ref{observation:exists-good-tuple}, we know that there exists an annotated tuple $(\emptyset,X',\pi,\gamma)$ that respects $R^*$ and $\phi^*$. Thus invoking Lemma~\ref{lemma:annotated-d-hs} on $(\emptyset,X',\pi,\gamma)$ returns a solution $S'$ to $(U',\capa',\cA')$ of size at most $\frac{4}{3}k$ satisfying  $|S'\cap X'_i|\leq 2$ for each $i\in [k]$. Further let $S$ be the set constructed from $S$ and returned by the algorithm.
Now the filter $X'$ is doing a standard bucketing technique for approximating weights. This combined with the fact that $|S'\cap X'_i|\leq 2$ gives us an error of at most $(2+\epsilon) W^*$. That is $W(S)\leq 2\epsilon W^*$ and $|S|\leq k$.

We now analyse the runtime of the algorithm. The time required to construct $\mathcal{F}$, which has size ${e^k k^{\mathcal{O}(\log k)}} \log n$ is  ${e^k k^{\mathcal{O}(\log k)}} n\log n$ ~\cite{alon1994color,DBLP:conf/focs/NaorSS95}. Let $n=|U|$ and $m:=|\cA|$. 
%Each call to Lemma~\ref{lemma:annotated-d-hs} runs in time $ $ and the Lemma is invoked $|\mathcal{F}|$ many times, thus the total runtime is $ $

\begin{enumerate}
    \item The first and second steps can be performed in time  $(n+m)^{\cO(1)}$.
    \item We obtain the family $\cF$ of size ${e^k k^{\mathcal{O}(\log k)}} \log n$ in ${e^k k^{\mathcal{O}(\log k)}} n\log n$ time in the third step.
    \item The number of possible functions $\eta$ is upper bounded by $\left(\frac{k \log n}{\epsilon}\right)^k \leq \left(\frac{k }{\epsilon}\right)^k +n$. Hence, we run Lemma~\ref{lemma:annotated-d-hs} on 
    $\left( \left(\frac{k }{\epsilon} \right)^k +n\right) \times e^k k^{\mathcal{O}(\log k)} \log n$ instances. Each such instance takes time $2^{k^{\cO(d)}}(n+m)^{\mathcal{O}(1)}$ to return the answer. Thus, the total running time of the algorithm is 
    $\left(\frac{k }{\epsilon} \right)^k 2^{k^{\cO(d)}}(n+m)^{\mathcal{O}(1)}$. 
    
\end{enumerate}
This concludes the proof. 
%\todo.
\end{proof}

\subsection{Proof of Lemma~\ref{lemma:annotated-d-hs}}
In this section we focus on the proof of our main technical lemma. 
We first introduce some additional notation required for our algorithm. See the overview (Section~\ref{sec:overview}) for an intuition of the definitions used in this section. 

\begin{definition}\label{definition:info-tuple}
    Given an annotated tuple $(S,X,\pi,\gamma)$ we define the information tuple of $(S,X,\pi,\gamma)$ as the tuple $(X'=X'_1\uplus\dots\uplus X'_{k-|S|},\mathsf{n},\mathsf{score})$ with:
    \begin{itemize}
          \item For each $i\in [k-|S|]$, $$X_i' := \{x \in X_i: \mathsf{cap}(x) \geq \sum_{s \in S} \gamma(i,s) \text{ and } |\mathcal{A}_E \odot x| \geq \gamma(i,E) , \text{ for each } E \in \cP_d(S)\}$$
    \item For $v\in X'$ with $v\in X'_i$,

    $$\mathsf{n}(v,E):=
    \min\bigg(
    \bigg\lceil 1+\frac{1}{3k} \bigg\rceil \cdot \gamma(i,E),
    |\cA_E \odot v|\bigg) \text{ for each $E\in \cP_d(S)$}$$
    
    $$\mathsf{n}(v,s):=\sum_{E\in \pi^{-1}(s)} \mathsf{n}(v,E) \text{ for each $s\in S$}$$
    
    $$\score(v,s):=\min\bigg(\mathsf{n}(v,s),\capa(v)-\sum_{s'\in S, s'\neq s} \gamma(i,s)\bigg) \text{ for each $s\in S$}$$
    \end{itemize}
\end{definition}

% Since $X'$ is just filtering out the elements in $X$ that can satisfy the coverage requirement according to the estimate $\gamma$, we have the following observation.
Assume $(S,X,\pi,\gamma)$ respects $S^*$ and $\phi^*$. Let $(X',\mathsf{n},\score)$ be the information tuple of $(S,X,\pi,\gamma)$. We observe that $X'$ filters out all vertices in $X$ that do not respect the estimate of coverage into each $E\in \cP_d$ according to $\gamma$. Thus $(S,X',\pi,\gamma)$ also respects $S^*$ and $\phi^*$-- Recall that $(S,X',\pi,\gamma)$ respects $S^*$. All properties except $2$ in Definition~\ref{definition:good-tuple} depend only on $S,\gamma,\pi$ and so hold directly for $(S,X',\pi,\gamma)$ as well. By how we defined $X'_i$ and using properties $3-5$ for $(S,X,\pi,\gamma)$, we can infer that $S^*\subseteq X'$ and $|X'_i\cap S^*|=1$ for each $i\in [k-|S|]$ proving property $2$ for $(S,X',\pi,\gamma)$. Thus we have the following observation.
 
\begin{observation}\label{observation:X'-good}
If $(S,X,\pi,\gamma)$ respects a solution $S^*$ and an assignment $\phi^*$, then $(S,X',\pi,\gamma)$ also respects $S^*$ and $\phi^*$.    
\end{observation}

\begin{definition}
    An extended annotated tuple is a tuple $(S,X,\pi,\gamma,\tau_1,\tau_2)$ such that:
    \begin{enumerate}
        \item $(S,X,\pi,\gamma)$ is an annotated tuple
        \item $\tau_1,\tau_2:S\rightarrow [k-|S|]$
    \end{enumerate}
\end{definition}

\begin{definition}\label{definition:candidate-set}
    Given an extended annotated tuple $(S,X,\pi,\gamma,\tau_1,\tau_2)$, we define the candidate set of $(S,X,\pi,\gamma,\tau_1,\tau_2)$ to be the subset $X'':=X''_1\uplus \cdots \uplus X''_{k-|S|}$ of $X$ such that:
    \begin{itemize}
        \item  For each $i\in [k-|S|]$, let $T(i,s)$ be the top $ dk^{10}$ vertices of $X'_{i}$ when sorted in non-increasing order of $\score(v,s)$. If $|X'_i|\leq dk^{11}$ let $X''_i:=X'_i$ and if not $X''_i:=\bigcup_{s\in \tau_{1}^{-1}(i)}T(i,s)$.
       % \todo{fix $f'(x)$}
    \end{itemize}
\end{definition}

\begin{definition}
    An extended annotated tuple $(S,X,\pi,\gamma,\tau_1,\tau_2)$ is good if there exists a solution $S^*$ to $\capinstance$ and an assignment $\phi^*:\cA\rightarrow S^*$ such that:
    \begin{enumerate}
        \item $(S,X,\pi,\gamma)$ is good and respects $S^*$ and $\phi^*$.
        %\item $S^*\cap X''=\emptyset$, where $X''$ is the candidate set of $(S,X,\pi,\gamma,\tau_1,\tau_2)$.
        \item For each $s\in S$, $\tau_1(s)=\max_{i\in [k-|S|]}\cov_{\phi^*}(v^*_i,\cA_s)$, where $v^*_i:=S^*\cap X_i$.
        \item For each $s\in S$, $\tau_2(s)=\max_{i\in [k-|S|]\setminus \tau_1(s)}\cov_{\phi^*}(v^*_i,\cA_s)$, where $v^*_i:=S^*\cap X_i$.
    \end{enumerate}
    We also say $(S,X,\pi,\gamma,\tau_1,\tau_2)$ respects $S^*$ and $\phi^*$.
\end{definition}
%\todo{remove $S^*\cap X$}
\begin{lemma}
\label{lemma:extended-annotated-tuple}
There exists an algorithm that takes as input an instance $\capinstance$, an annotated tuple $(S,X,\pi,\gamma,\tau_1,\tau_2)$ with candidate set $X''$ and in time $2^k\cdot (n+m)^{\cO(1)}$ either returns $\mathsf{FAIL}$ or succeeds and returns a solution $R$ to $\capinstance$ of size at most $\frac{4}{3}k$ such that $S\subseteq R$ and $|R\cap X_i|\leq 2$ for each $i\in [k-|S|]$. Furthermore if $(S,X,\pi,\gamma,\tau_1,\tau_2)$ respects a solution $S^*$ of size at most $k$ and an assignment $\phi^*$ and $S^*\cap X''=\emptyset$, it succeeds.
\end{lemma}
%\todo{change no to fail}
% \begin{definition}
%    Given sets $S,X\subseteq U$ with $X\cap S=\emptyset$, let $\mathcal{F}_{S,X}$ denote the set of all possible annotated tuples $(S,X,\pi,\gamma)$.  
% \end{definition}
We now provide our algorithm for Lemma~\ref{lemma:annotated-d-hs} using Lemma~\ref{lemma:extended-annotated-tuple} that we prove later. \\ \\
\textbf{Input:} An annotated tuple $(S,X,\pi,\gamma)$ and an instance $\capinstance$ \\
\textbf{Output:} $\mathsf{FAIL}$ or a solution to $\capinstance$ of size at most $\frac{4}{3}k$
\begin{enumerate}
    \item If $|S|=k$ and $S$ is a solution, return $S$ else return $\mathsf{FAIL}$.
    % \item For each $i\in [k-|S|]$ define
    % $$X_i' \leftarrow \{x \in X_i: \mathsf{cap}(x) \geq \sum_{s \in S} \gamma(i,s) \text{ and } |\mathcal{A}_s \odot x| \geq \gamma(i,s) \text{ for all } s \in S\}$$
    % \item For $i\in [k-|S|]$, $v\in X'_i$ and $s\in S$ define 
    
    % $$\mathsf{n}(v,s):=\sum_{E\in \pi^{-1}(s)}
    % \min\bigg(
    % \bigg\lceil 1+\frac{1}{k} \bigg\rceil \cdot \gamma(i,E),
    % |\cA_E \odot v|\bigg)$$
    
    % $$\score(v,s):=\min\bigg(\mathsf{n}(v,s),\capa(v)-\sum_{s'\in S, s'\neq s} \gamma(i,s)\bigg)$$
    % $$\score(v,s):=\min\{|\cA_s \odot v|,\capa(v)-\sum_{s'\in S, s'\neq s} \gamma(i,s)\}$$
    
    %\item For each $i\in [k-|S|]$, let $T(i,s)$ be the top $f(k)$ vertices of $X'_{i}$ when sorted in non-increasing order of $\score(v,s)$.
    \item For each pair of distinct functions $\tau_1: S\rightarrow [k-|S|]$ and $\tau_2:S\rightarrow [k-|S|]$ do:
    \begin{enumerate}
        % \item For each $i\in [k-|S|]$, if $|X'_i|\leq g(k)$  define $X''_i:=X'_i$ and if not $X''_i:=\bigcup_{s\in \tau_{1}^{-1}(i)}T(i,s)$ 
        % \item Define $X'':=\bigcup_{i\in [k-|S|]} X''_i$
        \item Construct the candidate set $X''$ for the extended annotated tuple $(S,X,\pi,\gamma,\tau_1,\tau_2)$.
        \item For each $i\in [k-1]$ and for each $v\in X''_i$
        \begin{itemize}
            \item Let $S':=S\cup \{v\}$ and $Y:=X\setminus X_i$
            \item For each annotated tuple $(S',Y,\pi',\gamma')\in \cF_{S',Y}$, apply the algorithm recursively on $(S',Y,\pi',\gamma')$. If the algorithm returns a set of size at most $\frac{4}{3}k$ return it.
        \end{itemize}
    \item Run the algorithm in Lemma~\ref{lemma:extended-annotated-tuple} on the extended annotated tuple $(S,X,\pi,\gamma,\tau_1,\tau_2,X'')$. If the algorithm returns a solution $R$, then output $R$.
    \end{enumerate}
    \item If no solution was output in the previous steps then return $\mathsf{FAIL}$
\end{enumerate}

It is clear that if the algorithm returns a set it is a solution of size at most $\frac{4}{3}k$ by Lemma~\ref{lemma:extended-annotated-tuple}. We now need to show that if $(S,X,\pi,\gamma)$ is good then the algorithm returns a solution of size at most $\frac{4}{3}k$.

From now, we fix $S^*$ to be a solution of size at most $k$ and $\phi^*$ to be an an assignment such that $(S,X,\pi,\gamma)$ respects $S^*$ and $\phi^*$. We prove that the algorithm returns a solution of size at most $\frac{4}{3}k$ by induction on the size of $S$. Base Case $|S|=k$: In this case $S\subseteq S^*$ since $(S,X,\pi,\gamma)$ respects $S^*$. Further since $|S^*|\leq k$, $S^*=S$ and thus the algorithm returns $S$.

We now assume that the algorithm returns a solution of size at most $\frac{4}{3}k$ for all $S$ of size greater than $\ell$. We prove the claim for $|S|=\ell$. 
% Let $X'=\bigcup_{i\in [k-|S|]} X'_i$. We observe that $X'$ filters out all vertices in $X$ that do not respect the estimate of coverage into each $E\in \cP_d$ according to $\gamma$. Formally, $(S,X',\pi,\gamma)$ is also good with respect to $S^*$ -- Recall that $(S,X',\pi,\gamma)$ respects $S^*$. Properties $1-2$ and $4-6$ in Definition~\ref{definition:good-tuple} depend only on $S,\gamma,\pi$ and so hold directly for $(S,X',\pi,\gamma)$ as well. By how we defined $X'_i$ and using properties $3-5$ for $(S,X,\pi,\gamma)$, we can infer that $S^*\subseteq X'$ and $|X'_i\cap S^*|=1$ for each $i\in [k-|S|]$ proving property $2$ for $(S,X',\pi,\gamma)$. 
We fix $\tau_1$ and $\tau_2$ according to $S^*$. Let  $v^*_i:= S^*\cap X'_i$ for each $i\in [k-|S|]$. Then for each $s\in S$ we set
\begin{equation}
\label{equation:tau}
    \begin{aligned}
        \tau_1(s)&:=\max_{i\in[k-|S|]} |\phi^{-1}(v^*_i)\cap \cA_s|\\
        \tau_2(s)&:=\max_{i\in[k-|S|]\setminus \tau_1(s)} |\phi^{-1}(v^*_i)\cap \cA_s|
    \end{aligned}
\end{equation}
% $$v^*_i:=S^*\cap X'_i$$
% $$\tau_1(s):=\max_{i\in[k-|S|]} |\phi^{-1}(v^*_i)\cap \cA_s|$$ $$\tau_2(s):=\max_{i\in[k-|S|]\setminus \tau_1(s)} |\phi^{-1}(v^*_i)\cap \cA_s|$$
%\textcolor{red}{Pick right $\tau_1$ and $\tau_2$ here}

%Next we fix $X''_i$ for each $i\in [k-|S|]$ to be the one in the iteration of the algorithm with $\tau_1$ and $\tau_2$. %Let $X'':=\bigcup_{i\in [k-s]}X''_i$. 

Let $X''$ be the candidate set of $(S,X,\pi,\gamma,\tau_1,\tau_2)$. We divide our proof into two cases: $(i)$ $S^*\cap X''\neq \emptyset$ and $(ii)$ $S^*\cap X''=\emptyset$.

We first prove case $(i)$. Let $v\in S^*\cap X''$ and $v\in X''_i$. We define $S':=S^*\cup\{v\}$ and $Y:=X\setminus X_i$. By Observation~\ref{observation:exists-good-tuple}, there exists an annotated tuple $(S',Y,\pi',\gamma')$ that is good with respect to $S^*$. The algorithm goes over every annotated tuple in $\cF_{S',Y}$ and thus calls the algorithm recursively on $(S',Y,\pi',\gamma')$. By induction since $|S'|>|S|$, this recursive call returns a solution of size at most $\frac{4}{3}k$.
%
%To complete our proof, we need to prove the claim for case $(ii)$. 

Case $(ii)$: Recall that $(S,X,\pi,\gamma)$ respects $S^*$ and $\phi^*$. By our choice of $\tau_1$ and $\tau_2$ and the fact that $S^*\cap X''=\emptyset$, the extended annotated tuple $(S,X,\pi,\gamma,\tau_1,\tau_2)$ respects $S^*$ and $\phi^*$.
Thus the algorithm runs Lemma~\ref{lemma:extended-annotated-tuple} with $(S,X,\pi,\gamma,\tau_1,\tau_2)$ which returns a solution of size at most $\frac{4}{3}k$.

We now discuss the running time. Step 1 in $(n+m)^{\mathcal{O}(1)}$ time. Constructing the candidate set also runs in $(n+m)^{\mathcal{O}(1)}$ time. 
Step $2$ makes the algorithm recursive with a branching factor of $e^k \times k^{k^d} \times \lceil\log_{1+1/3k}m\rceil^{k^{d+1}}$, (see Equation~\ref{equation:annotate-tuple-bound} for an explanation) and depth at most $k$. This leads to an algorithm with running time $2^{k^{\cO(d)}}(n+m)^{\mathcal{O}(1)}$. 

% equation:annotate-tuple-bound}
% \begin{aligned}
%$
%\leq k^{\cO(k^{d+1})}+m$$

\subsection{Proof of Lemma~\ref{lemma:extended-annotated-tuple}}

We first introduce the notion of independence that we will need. For this we recall Definition~\ref{definition:A_E-A_s} - Given a set $S\subseteq U$ and $\pi:S\rightarrow \cP_d(S)$, for each $s\in S$, $\cA_s:=\{A\in \cA_E: \pi(E)=s, E\in \cP_d(S)\}$.
\begin{definition}
    Given a set $S$ and function $\pi:\cP_d(S)\rightarrow S$, two elements $x,y\in U$ are said to be $(S,\pi,\rho)$-independent with respect to $S$ if for each $s\in S$, $|\cA_s\odot x \cap \cA_s\odot y| \leq \rho\min(|\cA_s\odot x|,|\cA_s\odot y|)$. Two elements that are not $(S,\pi,\rho)$-independent are said to be a conflicting pair.
\end{definition}

This is similar to the notion of conflict graph defined in Section~\ref{sec:overview}. We now state and prove a lemma that our algorithm will use as a subroutine.
% \begin{lemma}
% Given a set $S$, a function $\pi:S\rightarrow \cP_d(S)$, and a set $Y:=Y_1\uplus \cdots \uplus Y_{\ell}$ such that  $Y\subseteq U$, $\ell\leq k$ and $|Y_i|\geq f'(k)$ for each $i\in [\ell]$, there exists an algorithm that finds a $(S,\pi,\rho)$-independent subset $I$ of $Y$ such that for each $i\in [\ell]$, $|I\cap Y_i|\geq 2$ in time ... \todo{time and f'(k)}
% \end{lemma}
We first prove a claim that we will use.
\begin{claim} \label{inddegree}
For any $X\subseteq U$, there are at most $|X|dk/\rho$ many conflicting pairs in $X$.
\end{claim}
\begin{proof}
Let $x\in U$ be an element such that at least $1+dk/\rho$ elements, $\{y_1,y_2,\ldots y_{1+dk/\rho}\}\in X$  are in conflict with $x$ such that there $\exists s \in S$ with $|\cA_s\odot x \cap \cA_s\odot y_i| \leq \rho\min(|\cA_s\odot x|,|\cA_s\odot y_i|)$ and $|\cA_s\odot x|\leq |\cA_s\odot y_i|$. By pigeonhole principle, there exists an $s\in S$ such that there are $1+d/\rho$ many $y_i$s (say $Y'$)
     with  $|\cA_s\odot x \cap \cA_s\odot y_i| \leq \rho|\cA_s\odot x|$. 
      Since $\sum_{y\in Y'} |\mathcal{A}_s\odot y|\cap |\mathcal{A}_s\odot x|>(d/\rho)\cdot(\rho\cdot |\mathcal{A}_s\odot x|)= d \cdot |\mathcal{A}_s\odot x|$, by pigeonhole principle, there exists a set in $|\mathcal{A}_s\odot x|$ that is present in $|A\odot y|$ for $d+1$ many distinct $y$s in $Y'$, a contradiction to the fact that any set has at most $d$ elements. But this implies the existence of at most $|X|dk/\rho$ many conflicting pairs in $X$.
\end{proof}
\begin{lemma}\label{lemma:rho-ind-alg}
Given a set $S$, a function $\pi:S\rightarrow \cP_d(S)$, and a set $Y:=Y_1\uplus \cdots \uplus Y_{\ell}$ such that  $Y\subseteq U$, $\ell\leq k$ and $|Y_i|\geq dk^{10}$ for each $i\in [\ell]$, there exists an algorithm that finds a $(S,\pi,\rho)$-independent subset $I$ of $Y$ such that for each $i\in [\ell]$, $|I\cap Y_i|\geq 2$. For $\rho=1/k^4$ and $\ell\leq k$ it runs in time $(dk)^{\mathcal{O}(k)}$. 
\end{lemma}
\begin{proof}
We demonstrate our proof through the inductive construction of a set $I$. Note that in any $Y$
there are at least $|Y|(1-1/2k)$ elements that each have conflict with at most $(k+1)dk/\rho$ elements in $Y$. We assume all  $Y_is$ are of equal sizes (exactly $dk^{10}$) and hence every $Y_i$ has at least  $Y_i/2$ (half of its) elements that each have conflict with at most $(k+1)dk/\rho$ other elements in $Y$. We denote these \emph{low-degree} vertices in each $Y_i$ with $Y_i'$ where $|Y_i'|\geq |Y_i|/2 \geq (2\ell+1) (k+1)dk/\rho$.

With this we are ready to present a proof with an inductive construction of a required independent set $I$ in which every $Y_i'$ has two vertices. Note that we use independence in place of $(S,\pi,\rho)$-independence. We initialize $I_0=\emptyset$. At the $i$th step where $i\in[\ell]$, we select 2 vertices from $Y_i'$ that are independent among themselves as well as are independent with all the vertices in $I_{i-1}$. These vertices are then added to $I_{i-1}$ to form $I_i$, and ultimately, we define $I=I_{\ell}$. This is feasible due to the following facts. Due to Claim \ref{inddegree}, every vertex in $Y_i'$ has conflict with at most $(k+1)dk/\rho$ vertices in $Y$ and $|I_{i-1}|\leq 2\ell$. Hence in any $Y_i'$ there are at most $2(\ell-1) \cdot (k+1)dk/\rho$ many vertices that are in conflict with some vertex in $O_{i-1}$. And every $Y_i'$ containing at least $(2\ell+1)\cdot d/\rho +1$ vertices, has at least two vertices that are independent with all vertices in $I_{i-1}$.
\end{proof}

% \textcolor{blue}{
% We demonstrate our proof through the inductive construction of the set $I$. Note that in any $Y_i$
% there are at least $|Y_i|/2$ elements that each have confict with at most 
% In the proof of this lemma, we use independence in place of $(S,\pi,\rho)$-independence. We initialize $I_0=\emptyset$. At the $i$th step where $i\in[\ell]$, we select 2 vertices from $Y_i$ that are independent among themselves as well as are independent with all the vertices in $I_{i-1}$. These vertices are then added to $I_{i-1}$ to form $I_i$, and ultimately, we define $I=I_{\ell}$. This is feasible due to the following facts. Due to Claim \ref{inddegree}, every vertex in $U$ is non-independent with at most $dk/\rho$ vertices and $|O_{i-1}|\leq 2(k-|S|)$. Hence in any $Y_i$ there are at most $2(k-|S|-1) \cdot d/\rho$ many vertices that are non-independent with some vertex in $O_{i-1}$. And every $Y_i$ containing at least $2(k-|S|)\cdot d/\rho +1$ vertices, has at least two vertices that are independent with all vertices in $I_i$ }

%
We now state our algorithm for Lemma~\ref{lemma:extended-annotated-tuple} below.
\\ \\
\textbf{Input:} An extended annotated tuple $(S,X,\pi,\gamma,\tau_1,\tau_2)$ and an instance $\capinstance$ 
\\
\textbf{Output:} $\mathsf{FAIL}$ or a solution to $\capinstance$ of size at most $\frac{4}{3}k$

\begin{enumerate}
\item If for some $s\in S$, $\tau_1(s)=\tau_2(s)$ return $\mathsf{FAIL}$
\item  Construct the bipartite graph $H:=(S\uplus [k-s],E)$ where $E:=E_1\cup E_2$, $E_1:=\{(s,\tau_1(s)):s\in S\}$ and $E_2:=\{(s,\tau_2(s)):s\in S\}$.
\item Find a minimum size set $D\subseteq [k-|S|]$ that dominates $S$ in $H$ such that $\{i\in [k-|S|]:|\tau_1^{-1}(i)|\geq 2\} \subseteq D$. 
\item Construct the candidate set $X''$ of $(S,X,\pi,\gamma,\tau_1,\tau_2)$.
\item Construct a $(S,\pi,1/k^4)$-independent set $R\subseteq X''$ using Lemma~\ref{lemma:rho-ind-alg} on $S,\pi,X''$ such that:
    \begin{enumerate}
        \item $|R\cap X''_i|= 1$ for each $i\in [k-|S|]\setminus D$
        \item $|R\cap X''_i|= 2$ for each $i\in D$
        %\item $|R\cap X''_i|\geq 2$ for each $i\in [k-|S|]$ such that $|\tau^{-1}(i)|\geq 2$.
        %\item $R$ is $\frac{1}{k^2}$-conflict independent\todo{need stronger for bounds}
    \end{enumerate}
    %\item Return $R\cup S$ if it is a solution of size at most $\frac{4}{3}k$ else return $\mathsf{FAIL}$.
    \item Return $R\cup S$ if it is a solution of size at most $\frac{4}{3}k$ else return $\mathsf{FAIL}$.
\end{enumerate}
%\todo{runtime updated}
We first analyze our running time.  Given an extended annotated tuple $(S,X,\pi,\gamma,\tau_1,\tau_2)$ and an instance $\capinstance$, the algorithm runs in time $\underbrace{\mathcal{O}(2^k)}_\text{Step 3} + \underbrace{(n+m)^{\cO(1)}}_\text{Steps 1,2,4,6}+ \underbrace{dk^{\mathcal{O}(k)}}_\text{Step 5} \leq dk^{\mathcal{O}(k)}(n+m)^{\cO(1)}$ time.
%We bound the number of extended annotated tuples $(S,X,\pi,\gamma,\tau_1,\tau_2)$ by $\text{Choices of $S$} \cdot e^kk^{\mathcal{O}(k\log k)}\cdot k^{2^k} \cdot {\log n}^{k2^k} \cdot k^4$.\underbrace{e^k}_\text{steps 2-3}}

We first show that if the algorithm does not $\mathsf{FAIL}$ in step 1 and constructs the set $R\cup S$ then the set has size at most $\frac{4}{3}k$.
\begin{lemma}
If for each $s\in S$, $\tau_1(s)\neq \tau_2(s)$, then $|S\cup R|\leq \frac{4}{3}k$    
\end{lemma}
\begin{proof} 
 Note that $|S\cup R|=|S|+k-|S|+|D|=k+|D|$. We prove an upper-bound of $\frac{k}{3}$ on the size of a minimum dominating set $D$ of $H$ by considering the following two exhaustive cases. 
\begin{itemize}
    \item \textbf{Case 1 ($k-|S|<2|S|$)}: We construct a dominating set $D'\subseteq [k-|S|]$ such that $\{i\in [k-|S|]:|\tau_1^{-1}(i)|\geq 2\} \subseteq D'$ as follows. Initialize $H':=H$.
    For each $i\in [k-|S|]$ such that $|\tau^{-1}(i)|\geq 2$ (say $x$ many), we include $i$ into $D'$ and delete the vertices in $S$ that are already dominated, i.e, we delete $N_{H'}[i]$. With the inclusion of every such $i$, size of $H'$ decreases by 3 since $|N_{H'}[i]|\geq 3$ as $i$ dominates at least two vertices in $S$. After an exhaustive application of this inclusion procedure $|V(H')|\leq s-2x+k-x=k-3x$. 
    
    Observe that each vertex $s\in S$ in $H'$ has degree $2$ since $\tau_1(s)\neq \tau_2(s)$ for each $s\in S$. Thus applying Lemma \ref{domset} on $H'$ implies an existence of a dominating set $D''$ of size at most $1/3(k-3x)$ that can be obtained in $\mathcal{O}^*(2^k)$ time. Add such a $D''$ to $D'$. Therefore, $|D| \leq |D'| \leq |D'\setminus D''|+ |D''|\leq x + 1/3(k-3x) \leq \frac{k}{3}$.
    
    \item \textbf{Case 2 ($k-|S|\geq 2|S|$)}:  Note that there is a dominating set $D'=\{\tau_1(s)|s\in S\}$ dominating $S$ in $H$. Hence,  $|D| \leq |D'| \leq |s| \leq \frac{k}{3}$ (since $k-s\geq 2s\implies s\leq k/3$).
\end{itemize}
Thus $|S\cup R|\leq k+|D|\leq \frac{4k}{3}$ completing our proof.  
\end{proof}

It is clear that if the algorithm outputs the set $S\cup R$, then it is a solution of size at most $\frac{4}{3}k$. Further if $(S,X,\pi,\gamma,\tau_1,\tau_2)$ respects $S^*$, $\phi^*$ then by definition for each $s\in S$, $\tau_1(s)\neq \tau_2(s)$. So in this case, the algorithm will not fail in step 1, and will construct a set $S\cup R$ of size at most $\frac{4}{3}$ by our previous Lemma. So if we prove that $S\cup R$ is a solution when $S^*\cap X''=\emptyset$, we are done proving Lemma~\ref{lemma:extended-annotated-tuple}. We state this as a lemma below and devote the remainder of the section to prove it - this forms the crux of our proof.

% $(S,X,\pi,\gamma,\tau_1,\tau_2)$ is good, then $R\cup S$ is a solution.
% %
% Let $S^*$ be a solution of size at most $k$ and $\phi^*$ an assignment such that $(S,X,\pi,\gamma,\tau_1,\tau_2)$ respects $S^*$ and $\phi^*$ and let $X''$ be the candidate set of $(S,X,\pi,\gamma,\tau_1,\tau_2)$.
%
%We now state that $S\cup R$ is a solution.
\begin{lemma}\label{lemma:soln-SunionR}
If $(S,X,\pi,\gamma,\tau_1,\tau_2)$ respects a solution $S^*$ and an assignment $\phi^*$  and $S^*\cap X''=\emptyset$, where $X''$ is the candidate set of $(S,X,\pi,\gamma,\tau_1,\tau_2)$, then $S\cup R$ is a solution to $\capinstance$.
\end{lemma}
\subsubsection{Proof of Lemma~\ref{lemma:soln-SunionR}}
{\em From now for the rest of this section we fix $S^*$, $\phi^*$, the information tuple $(X',\mathsf{n},\score)$ of $(S,X,\pi,\gamma)$ and the candidate set $X''\subseteq X$ of $(S,X,\pi,\gamma,\tau_1,\tau_2)$. Further we define and fix for each $i\in [k-|S|]$, $v^*_i:=S^*\cap X_i$.} 
We note that such a $v^*_i$ exists since $(S,X',\pi,\gamma)$ respects $S^*$ and $\phi^*$. In fact observe that $v^*_i\in X'_i$ by Observation~\ref{observation:X'-good}. 

Let $\phi^*:\cA\rightarrow S^*$ be an assignment of $\mathcal{A}$ to $S^*$. 
For the proof we construct an assignment $\phi:\cA \rightarrow S\cup R$ such that for each $x\in S\cup R$, $\cov_{\phi}(x,\cA)\leq \capa(x)$. We first describe how to construct a partial assignment to $R$.

% Recall that $R\subseteq X''$. Let $\hat{R}$ be a subset of $o\in R$ satisfying $|X''_i\cap R|=1$, where $o\in X''_i$. Observe that for $o\in \hat{R}$ with $o\in X''_i$, it holds that $|\tau_1^{-1}(i)|\leq 1$. This is because otherwise $i\in D$ and that implies $|X''_i\cap R|=2$ by construction of $R$.

Recall that $X'':=\bigcup_{i\in [k-|S|]} X''_i$ and $R\subseteq X''$. Let $\hat{R}$ be a subset of $r\in R$, $r\in X''_i$ satisfying $|X''_i\cap R|=1$ and $|\tau_1^{-1}(i)|= 1$. 
We define a family $Y_{r,s}$ of sets from $\cA_s$ for each $r\in R$ and $s\in S$. 

\begin{definition}[$Y_{r,s},Y_r,Y'_r$]
   For $r\in \hat{R}$, $r\in X''_i$ and $s=\tau_1^{-1}(i)$, let $Y_{r,s}\subseteq \cA_s$ be a set such that:
   \begin{itemize}
       \item $|Y_{r,s}|=\score(r,s)$
       \item For each $E\in \cA_s$, $|Y_{r,s}\cap \cA_E|\leq \mathsf{n}(r,E)$
   \end{itemize}
   For all other $r\in R$, $r\in X''_i$ and $s\in S$, we define $Y_{r,s}\subseteq \cA_s$ to be a set such that:
   \begin{itemize}
       \item $|Y_{r,s}|=\gamma(i,s)$
       \item For each $E\in \cA_s$, $|Y_{r,s}\cap \cA_E|= \gamma(i,E)$
   \end{itemize}
   We define $Y_r:=\bigcup_{s\in S}Y_{r,s}$ and $Y'_r:=Y_r\setminus \bigcup_{r'\neq r:r'\in R}Y_{r'}$
\end{definition}
For each subset $R'\subseteq R$, we define $Y_{R',s},Y'_{R'},Y_{R',s}$ and $Y'_{R',s}$ in the natural way of taking unions of corresponding $Y$s over elements in $R$. By the way we define $(X',\mathsf{n},\score)$, $Y$ and the fact that $(S,X,\pi,\gamma,\tau_1,\tau_2)$ respects $S^*$ and $\phi^*$. We can show the following properties.
\begin{observation}
    For each equivalence class $E\in \mathcal{P}_d(S)$ and $r\in R$ with $r\in X''_i$, let $Y_{r,E}:=Y_{r,s}\cap \mathcal{A}_E$ we have,
    $$(1+1/3k)\cov_{\phi^*}(v^*_i,E)\geq (1+1/3k)\gamma(i,E)\geq\mathsf{n}(r,E)\geq |Y_{r,E}|\geq |Y'_{r,E}|$$
    %\geq (1-1/k^3)|Y_{r,E}|\geq (1-1/2k)\gamma(i,E)$$
\end{observation}
\begin{observation}
    For each $s\in \mathcal{P}_d(S)$ and $r\in R$ with $r\in X''_i$, we have
    $$(1+1/k)\cov_{\phi^*}(v^*_i,s)\geq |Y_{r,s}|\geq |Y'_{r,s}|\geq (1-1/k^3)|Y_{r,s}|\geq (1-1/2k)\cov_{\phi^*}(v^*_i,s)$$
\end{observation}
        
\begin{definition}[Partial assignment $\phi$ for $R$]
    For each $r\in R$ and $A\in Y'_r$, define $\phi(A)=r$. 
\end{definition}

We now state two Lemmas that we will use to completely the construction of $\phi$ and to prove it is a valid assignment.

\begin{lemma} 
\label{lemma:Rmorethanstar}
    For each $s\in S$, $\cov_{\phi}(R,\cA_s)\geq \cov_{\phi^*}(S^*\setminus S,\cA_s)$
\end{lemma}
\begin{lemma}
\label{lemma:S-coverage-E}
    For each $E\in \cP_d(S)$, $\cov_{\phi}(R,E)<\cov_{\phi^*}(\pi(E),E)$
\end{lemma}

Before we prove the two Lemmas, we complete the definition of $\phi$ by showing how to assign the remaining sets in $\cA$ to $S$.

\begin{definition}
    For each $E\in \cP_d$, $E$ is good if $\cov_{\phi}(R,E)\geq \cov_{\phi^*}(S^*\setminus S,E)$ and $E$ is bad if $\cov_{\phi}(R,E)< \cov_{\phi^*}(S^*\setminus S,E)$.  
\end{definition}

For each $E\in \c(P)$ that is bad, let $Y_E$ be an arbitrary set of $|\cov_{\phi^*}(S^*\setminus S,E)|-|\cov_{\phi}(R,E)|$ sets in $\cA_E\setminus \phi^{-1}(R)$. We then define $\phi$ for these sets below

$$\text{For each $A\in Y_E$, define $\phi(A):=\pi(E)$}$$

Observe that such an assignment can only be done because of Lemma~\ref{lemma:S-coverage-E}.

Let $\cA'$ be the family of sets in $\cA$ that is unassigned so far. 
$$\text{For each $A\in \cA'$, we define $\phi(A):=\phi^*(A)$}$$

We now argue that for each $v\in U$, $\phi^{-1}(U)\leq \capa(v)$. For each $r\in R$ with $r\in X''_i$ and $s\in S$, observe that $\gamma(i,s)\leq \cov_{\phi^*}(v^*_i,s)$ and that $\score(v,s)\leq \capa(v)-\cup_{i\in [k-|S|]:i\neq i'}\gamma(i,s)$.

Observe that since $R\subseteq X'' \subseteq X'$, 
$$\text{For each $r\in R$, $\capa(r)\geq \sum_{s\in S}\gamma(i,s)$}$$

This is because $X'$ was the set of vertices with this property that were filtered and selected for our use. Thus, for $r\in R$ such that $r\notin \hat{R}$

$$\cov_{\phi}(r,\cA)=\sum_{s\in S}\gamma(i,s)\leq \capa(r)$$

 % For $r\in \hat{R}$, $r\in X''_i$ and $s=\tau_1^{-1}(i)$
 
Now for each $r\in \hat{R}$ with $r\in X''_i$, recall that $|\tau_1^{-1}(i)|=1$. Let $s_r=\tau_1^{-1}(i)$, then by definition of $\score$ and $\phi$,
$$\cov_{\phi}(r,\cA)=\sum_{s\in S\setminus s_r}\gamma(i,s) + \score(r,s_r)\leq \capa(r)$$

\begin{claim}
    For each $s\in S$, $\cov_{\phi}(s,\cA)\leq \capa(s)$.
\end{claim}
\begin{proof}
The following is easy to see
$$\capa(s)\geq \cov_{\phi^*}(s,\cA) = \sum_{E\in \cP_d(s)}\cov_{\phi^*}(s,\cA_E) = \sum_{E\in \pi^{-1}(s)}\cov_{\phi^*}(s,\cA_E) +
\sum_{E\notin \pi^{-1}(s)}\cov_{\phi^*}(s,\cA_E)$$

We know by construction directly that
$$ \sum_{E\notin \pi^{-1}(s)}\cov_{\phi}(s,\cA_E) \leq \sum_{E\notin \pi^{-1}(s)}\cov_{\phi^*}(s,\cA_E)$$ 

Therefore we are left to prove the following: $$\sum_{E\in \pi^{-1}(s)}\cov_{\phi}(s,\cA_E)\leq \sum_{E\in \pi^{-1}(s)}\cov_{\phi^*}(s,\cA_E)$$
% \begin{equation*}
%     \begin{aligned}
%         \sum_{E\in \pi^{-1}(s)}\cov_{\phi}(s,\cA_E) \leq \sum_{E\in \pi^{-1}(s): E \text{ is bad}}\cov_{\phi}(s,\cA_E) +
%     \sum_{E\in \pi^{-1}(s): E \text{ is good}}\cov_{\phi}(s,\cA_E)  
%    \end{aligned}
% \end{equation*}

If $E\in \pi^{-1}(s)$ and $E$ is good, then $\cov_{\phi^*}(S^*\setminus S,\cA_E)\leq \cov_{\phi}(R,\cA_E)$. By our assignment scheme,
$$ \cov_{\phi}(s,\cA_E) = \cov_{\phi^*}(s,\cA_E)-(\cov_{\phi}(R,\cA_E)-\cov_{\phi^*}(S^*\setminus S,\cA_E))$$

If $E\in \pi^{-1}(s)$ and $E$ is bad, then $\cov_{\phi^*}(S^*\setminus S,\cA_E)>\cov_{\phi}(R,\cA_E)$. By our assignment scheme,
$$ \cov_{\phi}(s,\cA_E) = \cov_{\phi^*}(s,\cA_E)+\cov_{\phi^*}(S^*\setminus S,\cA_E)-\cov_{\phi}(R,\cA_E)$$

We know from Lemma~\ref{lemma:Rmorethanstar} that
$$\cov_{\phi}(R,\cA_s)\geq \cov_{\phi^*}(S^*\setminus S,\cA_s)$$

The above inequality directly gives us:
$$\sum_{E\in \pi^{-1}(s)}\cov_{\phi}(s,\cA_E) 
\leq \sum_{E\in \pi^{-1}(s): E \text{ is bad}}\cov_{\phi^*}(s,\cA_E) + \sum_{E\in \pi^{-1}(s): E \text{ is good}}\cov_{\phi^*}(s,\cA_E) 
\leq \sum_{E\in \pi^{-1}(s)}\cov_{\phi^*}(s,\cA_E)$$
\end{proof}

%%%%%%%%%%%%%%%%%%%%%%%%%%%%%%%%%%%%%%%
%%%%%%%%%%%%%%%%%%%%%%%%%%%%%%%%%%%%%%%
%%%%%%%%%%%%%%%%%%%%%%%%%%%%%%%%%%%%%%%
%%%%%%%%%%%%%%%%%%%%%%%%%%%%%%%%%%%%%%%
We now provide the Proof of Lemma~\ref{lemma:Rmorethanstar} and Lemma~\ref{lemma:S-coverage-E} one after the other to complete the proof of Lemma~\ref{lemma:soln-SunionR}.

\begin{proof}[Proof of Lemma~\ref{lemma:Rmorethanstar}]

Let $\rho=\frac{1}{k^4}$. Recall that by construction, $s$ is either dominated by $\tau_1(s)$ or $\tau_2(s)$. We divide into two cases, case $(i)$: $s$ is dominated by $\tau_1(s)$ in $D$ and $(ii)$ $s$ is dominated by $\tau_2(s)$ in $D$.

Recall by properties of $(S,X,\pi,\gamma,\tau_1,\tau)$ with respect to $S^*$ and $\phi^*$, $\tau_1{s}$ is the $i$ such that $v^*_i$ contributes the most to $\cA(s)$ in $\phi^*$ among $S\setminus S^*=\{v^*_1,\cdots,v^*_{k-|S|}\}$ and $\tau_2(s)$ is the $i$ such that $v^*_i$ contributes the second most to $\cA(s)$ among $S\setminus S^*$.

We define $R_1$ to be a subset of $R$ of size $k-|S|$ containing exactly one element from $X''_i\cap R$. Recall that by construction $R$ contains at least one element from $X''_i$.

\textit{Case $(i)$ ($s$ is dominated by $\tau_1(s)$ in $D$):} Here $|R\cap X_{\tau_1(s)}|\geq 2$ by construction. Let $r:=(R\setminus R_1)\cap X_{\tau_1(s)}$. 
We know that the $Y'_r$s are disjoint. Further for each $r'\in R_1$ with $r'\in X''_i$, we have $|Y'_{r',s}|\geq (1-1/2k)\cov_{\phi^*}(v^*_i,\cA_s)$. Thus we obtain,
$$ \cov_{\phi}(R_1,\cA_s)\geq (1-1/2k)\cov_{\phi^*}(S^*\setminus S,\cA_s)$$
Next by definition of $\tau_1(s)$ we know that $\cov_{\phi^*}(v^*_{\tau_1(s)},\cA_s)\geq (1/k)\cov_{\phi^*}(S^*\setminus S,\cA_s)$. Thus we have, 
$$\cov_{\phi}(r,\cA_s) 
\geq (1-1/2k)\cov_{\phi^*}(v^*_i,\cA_s)
\geq(1/k)(1-1/2k)\cov_{\phi^*}(S\setminus S^*,\cA_s)$$

Combining these two equations we get,
\begin{equation*}
\begin{aligned}
    \cov_{\phi}(R_1\cup r,\cA_s)&\geq \cov_{\phi^*}(S\setminus S^*,\cA_s)\\
    => \cov_{\phi}(R,\cA_s)&\geq \cov_{\phi^*}(S\setminus S^*,\cA_s)\\
\end{aligned}
\end{equation*}

\textit{Case $(ii)$ ($s$ is dominated by $\tau_2(s)$ in $D$):} Here $|R\cap X_{\tau_2(s)}|\geq 2$ by construction. Let $r_1:=R_1\cap X_{\tau_1(s)}$ and let $r_2:=(R\setminus R_1)\cap X_{\tau_2(s)}$. 
Here we use that $\score(r_1,s)\geq cov_{\phi^*}(v^*_{\tau_1(s)},s)$ -- this is true by how $X''$ is constructed. Thus by definition $|Y_{r_1,s}|\geq \score(r_1,s)\geq cov_{\phi^*}(v^*_{\tau_1(s)},s)$. 
Further $|Y_{r_1,s}\cap \bigcup_{r\in R\setminus r_1} Y_{r,s}|\leq (1/k^3)\cov_{\phi^*}(S\setminus S^*\cup v^*_{\tau_1(s)},\cA_s)$ due to $R$ being $(S,\pi,1/k^4)$-bounded and the properties on $|Y_{r,s}|$. So we have:

$$\cov_{\phi}(r_1,\cA_s)=|Y'_{r_1,\cA_s}|\geq \cov_{\phi^*}( v^*_{\tau_1(s)},\cA_s) - (1/k^3)\cov_{\phi^*}(S\setminus S^*\cup v^*_{\tau_1(s)},\cA_s)$$
Next we observe that $v^*_{\tau_2(s)}$ is the top contributor to $\cA_s$ among all vertices in $S^*\setminus S\cup v^*_{\tau_1(s)}$. Thus we obtain the below two inequalities by using a similar argument as that of case (i).

\begin{equation}
\begin{aligned} 
    \cov_{\phi}(R_1\setminus r_1,\cA_s)&\geq (1-1/2k)\cov_{\phi^*}(S\setminus S^*\cup v^*_{\tau_1(s)},\cA_s)\\
    \cov_{\phi}(r_2,\cA_s)&\geq (1/k)(1-1/2k)\cov_{\phi^*}(S\setminus S^*\cup v^*_{\tau_1(s)},\cA_s)\\
\end{aligned}
\end{equation}
Combining all the three equations above we get,
\begin{equation}
\begin{aligned}
    \cov_{\phi}(R_1\cup r_2,\cA_s)&\geq \cov_{\phi^*}(S\setminus S^*,\cA_s)\\
    => \cov_{\phi}(R,\cA_s)&\geq \cov_{\phi^*}(S\setminus S^*,\cA_s)\\
\end{aligned}
\end{equation}
\end{proof}

\begin{proof}[Proof of Lemma~\ref{lemma:S-coverage-E}]
We first prove an useful claim.
    \begin{claim} 
\label{claim:E-coverage-S}
If $S^*\cap X''=\emptyset$, then for each $E\in \cP_d(S)$, $\cov_{\phi^*}(S,E)\geq (1-\frac{1}{k^2})|\cA_E|$
\end{claim}
\begin{proof}

Let $f'(k)=dk^{10}$ and $f(k)=k^2$ Let us assume the claim is false and that $\cov_{\phi^*}(S,E)< (1-\frac{1}{f(k)})|\cA_E|$. Recall that $(S,X,\pi,\gamma)$ respects $S^*$ and $\phi^*$ and that $v^*_i=S\cap X_i$.
Since $S\subseteq S^*$ and $|S|\leq k$, there exists an $i\in [k-|S|]$ such that, %there exists a $v^*_i\in S^*\setminus S$, where $v^*_i\in X_i$ such that 
$$\cov_{\phi^*}(v^*_i,E)\geq \frac{1}{k}\frac{1}{f(k)}|\cA_E|$$ 

Let $i\in [k-|S|]$ be one that satisfies the above inequality. Since $(S,X,\pi,\gamma)$ respects $S^*$ and $\phi^*$, $$\gamma(i,E)\geq \frac{1}{1+1/3k}\cov_{\phi^*}(v^*_i,E) \geq \frac{3k}{3k+1}\frac{1}{k}\frac{1}{f(k)}|\cA_E| = \frac{3}{(3k+1)f(k)}|\cA_E|$$

Let $X'$ be the one from the information tuple of $(S,X,\pi,\gamma)$. By Definition~\ref{definition:info-tuple} for each $x_i\in X''_i$ since $X''_i\subseteq X'_i$, 
$$|\cA_E \odot x_i|\geq \gamma(i,E) \geq \frac{3}{(3k+1)f(k)}|\cA_E|$$
Also since $X''$ is the candidate set of $(S,X,\pi,\gamma,\tau_1,\tau_2)$, by Definition~\ref{definition:candidate-set} and the fact that $S^*\cap X''=\emptyset$, we have:
$$|X''_i|\geq f'(k)$$ 
%\todo{f'(k) is the top f'(k) elements}

% Further note that for each $x_i\in X''_i$, $\score(x_i,s)\geq \score(v^*_i,s) \geq cov_{\phi^*}(v^*_i,s)\geq $.
%

Let $\eta$ be the number of incidences of $\cA$ in $X''_i$, that is $\eta:= \sum_{x_i\in X''_i}|\cA_E\odot x_i|$. Since each set $A\in \cA$ has $|A|\leq d$, $\eta\leq d|\cA_E|$.
But since for each $x_i\in X''_i$, since $|\cA_E \odot x_i|\geq \frac{3}{(3k+1)f(k)}|\cA_E|$ we have $$\eta\geq |X''_i|\frac{3}{(3k+1)f(k)}|\cA_E| \geq \frac{3f'(k)}{(3k+1)f(k)}|\cA_E|>d|\cA_E|$$
%\todo{use dependence of f(k), f'(k) to set parameter}
This contradicts the fact that $\eta\leq d|\cA_E|$. Therefore our assumption was incorrect and so $\cov_{\phi^*}(S,E)\geq (1-\frac{1}{f(k)})|\cA_E|$.
\end{proof}
For each $r\in R$ with $r\in X''_i$ and our observations, we know the following:
$$(1+1/3k)\cov_{\phi^*}(v^*_i,E)\geq (1+1/3k)\gamma(i,E)\geq\mathsf{n}(r,E)\geq |Y_{r,E}|\geq |Y'_{r,E}|$$

By our previous claim, $$\cov_{\phi^*}(S\setminus S^*,E)\geq (1-1/f(k))|E|$$

In addition using $\cov_{\phi}(r,E)=|Y'_{r,E}|\leq (1+1/3k)\cov_{\phi^*}(v^*_i,E)$ and $|R\cap X''_i|\leq 2$ gives
$$\cov_{\phi}(R,E)\leq 2(1+1/3k)\cov_{\phi^*}(S\setminus S^*,E)\leq 2(1+1/3k)(1/f(k))|E|$$
% Further by our definition of $\mathsf{n}$, 
% For each $r\in R$ and $E\in \cP_d$, $\mathsf{n}(r,E)\geq|Y'_{r,E}|$

Next by our selection of $\pi$, $\pi(E)$ covers the maximum from $S^*\setminus S$ and so $$\cov_{\phi^*}(\pi(E),E)\geq 1/k(1-1/f(k))|E|$$

The bounds on $\cov_{\phi^*}(\pi(E),E)$ and $\cov_{\phi^*}(R,E)$ shows $\cov_{\phi^*}(R,E)<\cov_{\phi^*}(\pi(E),E)$.
\end{proof}

\section{Parameterized Inapproximability of unweighted {{\sc Capacitated Vertex Cover}} with multi-edges}
\label{sec:hardness-unwt}
We know that {\sc Capacitated Vertex Cover} admits a FPT algorithm when the input graph is simple (that is, has no parallel edges). In this section, we show the parameterized inapproximability of unweighted {\sc Capacitated Vertex Cover} with multi-edges. The reduction happens in two steps, in the first we step we reduce from {\sc $3$-Regular $2$-CSP} to {\sc Multi-Dimensional Knapsack} and in the second step we reduce {\sc Multi-Dimensional Knapsack} to {\sc Capacitated Vertex Cover} with multi-edges. 
\subsection{Parameterized Inapproximability of {\sc $3$-Regular $2$-CSP}}\label{sec:constraintSatisfaction}

% Given a set of variables $X=\{x_1,x_2,\ldots,x_k\}$ and a family of domains ${\cal D}=\{D_1,D_2,\ldots,D_k\}$, a {\em binary constraint} is a pair $c=((x_i,x_j),R)$ where $x_i,x_j\in X$, $i\neq j$, and $R$ is a binary relation over $D_i\times D_j$. An {\em evaluation} is a function $\psi: X\rightarrow \bigcup{\cal D}$ such that for all $x_i\in X$, $\psi(x_i)\in D_i$. An evaluation $\psi$ is said to {\em satisfy} $((x_i,x_j),R)$ if $(\psi(x_i),\psi(x_j))\in R$. Moreover, given a multiset $C$ of binary constraints, an evaluation $\psi$ is said to {\em satisfy} $C$ if it satisfies every constraint $c\in C$. For all $i\in[k]$, let $C_i\subseteq C$ denote the sub(multi)set of constraints where $x_i$ occurs, and let $s_i=|C_i|$. We assume w.l.o.g.~that for all $i\in[k]$, $s_i\geq 1$, that is, for every variable in $X$, there exists at least one binary constraint where it occurs.
We first define our source problem. Toward that we first define constraint satisfaction problems (CSPs) of arity two (also called binary constraints).) We follow the notation and definitions of
the seminal paper of Guruswami et al.~\cite{DBLP:conf/stoc/GuruswamiLRS024}. 
Formally, a CSP instance $G$ is a quadruple $(V(G), E(G), \Sigma, \{C_{e}\}_{e\in E(G)})$, where:
\begin{itemize}
\setlength{\itemsep}{-2pt}
    \item $V(G)$ is for the set of variables.
    \item $E(G)$ is for the set of constraints.
    Each constraint $e=\cbra{u_e,v_e} \in E(G)$ has arity $2$ and is related to two distinct variables $u_e,v_e\in V(G)$.
    The \emph{constraint graph} is the undirected graph on the vertices $V(G)$ and the edges $E(G)$. Note that we allow multiple constraints between the same pair of variables and thus the constraint graph may have parallel edges.
    \item $\Sigma$ is for the alphabet of each variable in $V(G)$. We use $\Sigma=[n]$. 
    %For convenience, we sometimes have different alphabets for different variables and we will view them as a subset of a grand alphabet $\Sigma$ with some natural embedding.
    \item  Given a constraint $e\in E(G)$, $C_e\subseteq \Sigma \times \Sigma =[n]\times [n]$. Furthermore, given $\{C_{e}\}_{e\in E(G)}$, we can define $\{\Pi_{e}\}_{e\in E(G)}$, the set of validity functions of constraints.  
    Given a constraint $e\in E(G)$, the validity function $\Pi_e(\cdot,\cdot)\colon\Sigma\times\Sigma\to\bin$ checks whether the constraint $e$ between $u_e$ and $v_e$ is satisfied. That is, $\Pi_e(\cdot,\cdot)$ assigns $1$ if and only of the  tuple is in $C_e$. 
\end{itemize}
We use $|G|=(|V(G)|+|E(G)|)\cdot|\Sigma|$ to denote the \emph{size} of a CSP instance $G$.

\paragraph*{Assignment and Satisfaction Value.}
% An \emph{assignment} is a function $\sigma\colon V\to \Sigma$ that assigns each variable a value in the alphabet. 
% The \emph{satisfiability value} for an assignment $\sigma$, denoted by $\val(G,\sigma)$, is the fraction of constraints satisfied by $\sigma$, i.e.,
% $\val(G, \sigma)=\frac{1}{|E|}\sum_{e\in E} \Pi_e(\sigma(u_e),\sigma(v_e))$.
% The satisfiability value for $G$, denoted by $\val(G)$, is the maximum satisfiability value among all assignments, i.e., $\val(G) = \max_{\sigma\colon V\to \Sigma} \val(G, \sigma)$. 
% We say that an assignment $\sigma$ is a \emph{solution} to a CSP instance $G$ if $\val(G,\sigma) = 1$, and $G$ is \emph{satisfiable} iff $G$ has a solution. When the context is clear, we omt $\sigma$ in the description of a constraint, i.e., $\Pi_e(u_e,v_e)$ stands for $\Pi(\sigma(u_e),\sigma(v_e))$.

An \emph{assignment} is a function $\sigma\colon V(G)\to \Sigma$ that assigns to each variable a value from the alphabet. The \emph{satisfaction value} for a mapping $\sigma$, denoted as $\val(G,\sigma)$, represents the proportion of constraints that $\sigma$ satisfies, i.e., $$\val(G, \sigma)=\frac{1}{|E|}\sum_{e\in E} \Pi_e(\sigma(u_e),\sigma(v_e)).$$ 
The satisfaction value for $G$, denoted by $\val(G)$, is the highest satisfaction value across all mappings, i.e., $\val(G) = \max_{\sigma\colon V(G)\to \Sigma} \val(G, \sigma)$. We define an assignment $\sigma$ as a \emph{solution} to a CSP instance $G$ if $\val(G,\sigma) = 1$, and say $G$ is \emph{satisfiable} if and only if $G$ has a solution. When the context is clear, $\sigma$ is omitted in the constraint description; therefore, $\Pi_e(u_e,v_e)$ represents $\Pi(\sigma(u_e),\sigma(v_e))$.

We will be working with the {\sc $3$-Regular $2$-CSP} problem. An input to this problem consists of a 2CSP $G$ with $k$ variables over size-$n$ alphabets, where the constraint graph $G$ is $3$-regular. The question is whether $G$ is satisfiable. 

\begin{proposition}[\cite{DBLP:conf/stoc/GuruswamiLRS024,DBLP:conf/soda/LokshtanovR0Z20}]
\label{prop:ETHPIH}
%Unless ETH fails, there is an absolute constant
% \footnote{The exact constant here is not important. Starting from a constant $\eps>0$, one can boost it to $1-\eta$ for any constant $\eta>0$ by standard reductions.} 
% $\varepsilon>0$, such that no fixed parameter tractable algorithm which, takes as input a 2CSP $G$ with $k$ variables over size-$n$ alphabets of  {\sc $3$-Regular $2$-CSP}, can decide whether $G$ is satisfiable or at least $\varepsilon$ fraction of constraints must be violated.
Unless ETH fails, there is an absolute constant
$\varepsilon >0$, such that no algorithm that takes as input $(G, k)$  of {\sc $3$-Regular $2$-CSP} with $k$ variables over size-$n$ alphabets, runs in $f(k)n^{\cO(1)}$ time and distinguishes between instances $G$ of {\sc $3$-Regular $2$-CSP}, that have $\val(G)=1$ from instances $G$ having $\val(G)\leq 1-\varepsilon$. 
\end{proposition}

\subsection{Parameterized Inapproximability of {\sc Multi-Dimensional Knapsack}} 

\begin{comment}

\todo{in the instance notation instead of Cuv in E(G), use a better notation it is defined for all uv}
\begin{lemma}
\textcolor{red}{CSP inapproximability result here}
\end{lemma}
In this section, we demonstrate a \textcolor{red}{what is this called?}\emph{parameter/gap-preserving reduction} from an instance of the 3-regular 2-CSP problem to an instance of the {\sc Multi-Dimensional Knapsack} problem. We start by providing formal definitions of the two problems. In a 3-regular 2-CSP problem, we are given an instance of $(G, C_{(u,v)\in E(G)} \subseteq [n] \times [n],k)$, where $G$ is a 3-regular Gaifmann graph with $k$ edges, $[n]$ is the domain, and each edge $(u, v) \in E(G)$ has a corresponding constraint $C_{(u,v)} \subseteq [n] \times [n]$. An assignment for the CSP instance is a function $f : V(H) \rightarrow [n]$. An edge $(u, v)$ is said to be satisfied by an assignment $f$ if and only if $(f(u), f(v)) \in C_{(u,v)}$. The objective of the 3-regular 2-CSP problem is to determine if there exists an assignment $f$ that satisfies all the $k$ edges. Such an assignment is referred to as a satisfying assignment. 
\end{comment}

In this section, we give a gap-preserving reduction from {\sc $3$-Regular $2$-CSP} to {\sc Multi-Dimensional Knapsack}, which will show that the latter problem is hard to approximate in FPT time.  Recall that, in a {\sc Multi-Dimensional Knapsack} problem,  input consists of $n$ vectors 
%given an instance $(\mathbb{V}, \mathfrak{t}, k)$, where 
$\mathbb{V} = \{\mathfrak{v}_1, \ldots , \mathfrak{v}_n\}$, such that
each  $\mathfrak{v}_i \in \{0,1,\ldots,M\}^d$, a target vector $\mathfrak{t} \in \{0,1,\ldots,M\}^d$, and a positive integer $k$.  
The task is to determine whether there exists a subset $\mathbb{U} \subseteq \mathbb{V}$ of size $k$ such that $S(\mathbb{U})=\sum_{\mathfrak{v}_i\in \mathbb{U}} \mathfrak{v}_i  \geq \mathfrak{t}$? For two vectors $\mathfrak{a},\mathfrak{b}$, $\mathfrak{a}>\mathfrak{b}$ if and only if $a_i \geq b_i$ $\forall i \in[d]$ where $\mathfrak{a}=(a_1,\ldots a_d)$ and $ \mathfrak{b}=(b_1,\ldots,b_d)$.

% Note that the number of vertices (also referred to as variables) in $G$ is $2/3k$, since $3|V(G)|=2|E(G)|$ (sum of the degrees of all vertices). In a {\sc Multi-Dimensional Knapsack} problem, one is given an instance $(\mathbb{V}, \mathfrak{t}, k)$, where $\mathbb{V} = \{\mathfrak{v}_1, \ldots , \mathfrak{v}_n\}$, such that
% each $\mathfrak{v}_i \in \mathbb{V}$ is a $d$-dimensional vector, i.e., $\mathfrak{v}_i \in Z^
% d_{
% \geq0}$
% ; and $\mathfrak{t} \in Z^
% d_{
% \geq0}$
% is the $d$-dimensional
% target vector. The task is to determine whether there exists a subset $\mathbb{U} \subseteq \mathbb{V}$ of size $k$ such that $\sum_{\mathfrak{v}_i\in \mathbb{U}} \mathfrak{v}_i = S(\mathbb{U})\geq \mathfrak{t}$? For two vectors $\mathfrak{a},\mathfrak{b}$, $\mathfrak{a}>\mathfrak{b}$ if and only if $a_i \geq b_i$ $\forall i \in[d]$ where $\mathfrak{a}=(a_1,\ldots a_d)$ and $ \mathfrak{b}=(b_1,\ldots,b_d)$. 
\medskip 

\noindent\textbf{Construction:} 
% Given an instance $(G, C_{(u,v)\in E(G)} \subseteq [n] \times [n],k)$ of 3-regular 2-CSP problem, we construct an equivalent instance  $(\mathbb{V}, \mathfrak{t}, 5/3k)$ of {\sc Multi-Dimensional Knapsack} as follows. 
Given an instance $G=(V(G), E(G), \Sigma, \{C_{e}\}_{e\in E(G)})$ of {\sc $3$-Regular $2$-CSP}, we construct an instance  $(\mathbb{V}, \mathfrak{t}, 5/2k)$ of {\sc Multi-Dimensional Knapsack} as follows. We first define the number of dimensions each vector has and their significance.  The dimensions of the vectors in $\mathbb{V}$ are of the following two types.
\begin{itemize}
\setlength{\itemsep}{-1pt}
    \item There is a \emph{guard} dimension for every variable and for every edge. We will denote it by {\em $z$-guard dimension}, $z\in V(G)\cup E(G)$. 
    \item For every variable-edge incidence there are two dimensions, $v-e_{+}$ and $v-e_{-}$.  That is, if $v$ is incident to an edge $e$, then we have two dimensions that correspond to it, namely, $v-e_{+}$ and $v-e_{-}$. These will be called incidence dimensions. 
    \item Hence, the number of dimensions $d=\underbrace{|V(G)|+|E(G)|}_\text{guard dimensions}+\underbrace{4\cdot |E(G)|}_\text{incidence  dimensions}=8.5k$.  
\end{itemize}

Let $Q=10n$ be a {\em fixed number}. We add vectors to $\mathbb{V}$ as follows. 

\begin{itemize}
\setlength{\itemsep}{-1pt}
\item For every $x \in V(G)$ and $i\in[n]$, we add a vector $\mathfrak{v}_{(x,i)}$ which has $1$ in the $x$-guard dimension and for every edge $e$ incident on $x$, in the $x-e_+$ dimension it has value $Q+i$ and in the $x-e_-$ dimension has value $Q-i$. In every other dimension the vector has value $0$.

\item For every $(a,b)\in C_{(u,v)}$ (we denote $(u,v)$ as $e$), we add a vector ($\mathfrak{v}_{(a\leftarrow u, \text{ } b\leftarrow v})$) that has 1 in the $e$-guard dimension, $Q-a$ in $u-e_+$ dimension, $Q+a$ in $u-e_-$ dimension, $Q-b$ in $v-e_+$ dimension, and $Q+b$ in $v-e_-$ dimension. In every other dimension the vector has value $0$.
\end{itemize}

\begin{figure}[t]
\includegraphics[scale=1.0]{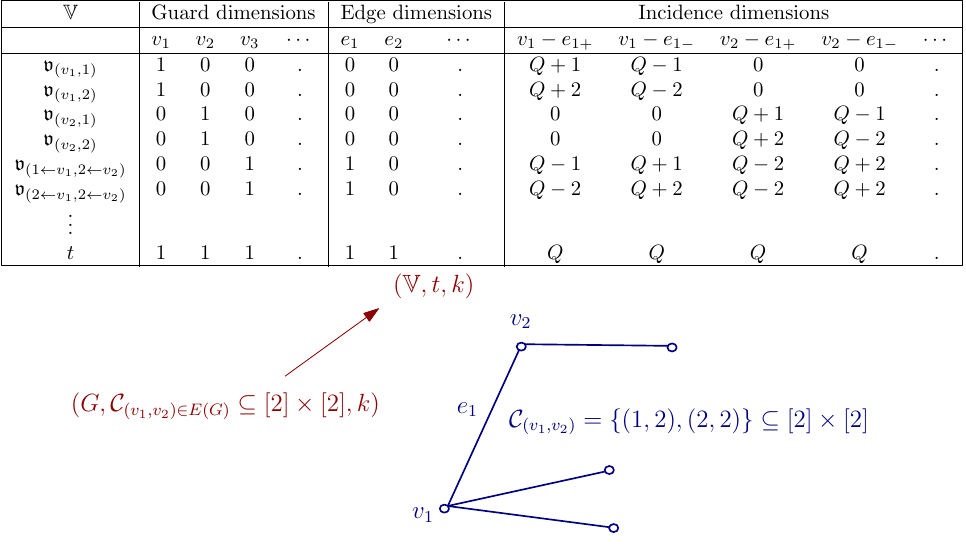}
\caption{An illustration of the construction.}
\label{fig:exConsCSPMDK}
\end{figure}

The target vector $\mathfrak{t}$ has 1 in all guard dimensions and $2Q$ in all incidence dimensions. This concludes the construction.  
See Figure~\ref{fig:exConsCSPMDK} for an illustration of the above construction. 
% In the diagram presented below, we outline the instance construction for a specific example. 

\begin{theorem} \label{cspeqv}
% Unless ETH fails, there is an absolute constant 
% $\delta>0$, such that no fixed parameter tractable algorithm which, takes as input $(\mathbb{V}, \mathfrak{t}, k)$ of {\sc Multi-Dimensional Knapsack}, can distinguish between instances that have  optimum solution of size at most $k$ instance or there is no solution of size at most $(1+\delta)k$.   
Unless ETH fails, there is an absolute constant
$\delta>0$, such that no algorithm that takes as input $(\mathbb{V}, \mathfrak{t}, k)$  of {\sc Multi-Dimensional Knapsack}, runs in $f(k,d)n^{\cO(1)}$ time and distinguishes between instances $(\mathbb{V}, \mathfrak{t})$ of {\sc Multi-Dimensional Knapsack} that have a solution of size at most $k$ from instances $(\mathbb{V}, \mathfrak{t})$ having no solution of size $(1+\delta)k$. The dimension of the input instance is at most $8.5k$.  
% Unless ETH fails, there is an absolute constant
% $\delta>0$, such that no algorithm that takes as input $(\mathbb{V}, \mathfrak{t}, k)$  of {\sc Multi-Dimensional Knapsack}, runs in $f(k,d)n^{\cO(1)}$ time and distinguishes between instances $(V,t)$ of {\sc Multi-Dimensional Knapsack} that have a solution of size at most $k$ from instances $(V,t)$ having no solution of size $(1+\delta)k$. The dimension of the input instance is at most $8.5k$. 

\end{theorem}
% % \textcolor{red}{Define opt of the instance. There is no algorithm that runs in time $F(k,d)$ time and distinguishes between instances such that $OPT<=k$ and that $OPT$ is $\leq (1+\delta)k$ For an instance $(V,T,K)$ and returns yes or no such that
% % }

% % \textcolor{red}{
% % There exists a constant such that there is no FPT algorithm that distinguishes between instances $(V,T,k)$ that have a solution of size at most k and no solution of size $(1+\delta)k$.
% % }

% \textcolor{red}{
% There exists no  algorithm that takes as input $(V,T,k)$ runs in FPT time and distinguishes between instances $(V,t)$ that have a solution of size at most $k$ from instances $(V,t)$ having no solution of size $(1+\delta)k$.
% }
\begin{proof}
%($\rightarrow$) 
We first show soundness followed by completeness. 

\noindent 
{\bf Soundness.} Let $G=(V(G), E(G), \Sigma, \{C_{e}\}_{e\in E(G)})$ be an instance of {\sc $3$-Regular $2$-CSP} such that $\val(G) = 1$. Furthermore, let $f$ be a solution assignment.  That is, $\val(G,f) = 1$. 
Let $\mathbb{U}$ be the collection of $2.5k$ vectors:
\begin{itemize}
    \item one created for each $x\in V(G)$ and $f(x)$, and one for each $(f(u),f(v))\in C_{(u,v)}$, where ${(u,v)\in E(G)}$. That is, $$\mathbb{U}=
    \Big\{\mathfrak{v}_{(x,f(x))}| x\in V(G)\Big\} \uplus \Big\{\mathfrak{v}_{(f(u)\leftarrow u, f(v)\leftarrow v)}| (u,v)\in E(G)\Big\}. $$
\end{itemize}

We claim $S(\mathbb{U})=\sum_{\mathfrak{v}_i\in \mathbb{U}} \mathfrak{v}_i  \geq \mathfrak{t}$. Note that in all guard dimensions ($8.5k$ many), $S(\mathbb{U})$ has value $1$  from construction. Let $x-e_+$ and $x-e_-$ be two dimensions that correspond to the edge $e=(x,y)$ and the vertex $x$ incident to $e$. In dimensions $x-e_-$ and $x-e_+$, the vector $\mathfrak{v}_{(x,f(x))}$ has values $Q-f(x)$ and $Q+f(x)$, respectively, while the vector $\mathfrak{v}_{(f(x)\leftarrow x, f(y)\leftarrow y)}$ has values $Q+f(x)$ and $Q-f(x)$, respectively. And every other vector in $\mathbb{U}$ has value 0 in these dimensions. Thus, in both of these dimensions $S(\mathbb{U})$ has the value $2Q$. Since $x$ is an arbitrary variable/vertex and $e$ being an arbitrary edge incident to $x$, it ensures the same argument can be made for every incidence dimension, making $\mathbb{U}$ a solution to the $(\mathbb{V},\mathfrak{t},8.5k)$ instance of {\sc Multi-Dimensional Knapsack}  problem. 

\medskip

\noindent 

\noindent 
{\bf Completeness.}  Let $G=(V(G), E(G), \Sigma, \{C_{e}\}_{e\in E(G)})$ be an instance of {\sc $3$-Regular $2$-CSP} such that $\val(G) \leq  1-\varepsilon$. Here, $\varepsilon>0$ is an absolute constant mentioned in Proposition~\ref{prop:ETHPIH}. In this case, we will show that there exists an absolute constant 
$\delta>0$, depending on $\varepsilon$, such that the reduced instance $(\mathbb{V},\mathfrak{t},2.5k)$ of {\sc Multi-Dimensional Knapsack} does not have a solution of size $(1+\delta)2.5k$. 

For a contradiction, assume that $(\mathbb{V},\mathfrak{t},2.5k)$ has a set of $(1+\delta)2.5k$ vectors, say $\mathbb{U}\subseteq \mathbb{V}$ such that $S(\mathbb{U})\geq \mathfrak{t}$. To achieve the target value of $1$ in all guard dimensions, we must have a subset $\mathbb{U}'\subseteq \mathbb{U}$ containing $2.5k$ vectors. In particular, it contains vectors $\mathfrak{v}_{(x,j)}$, $x\in V(G)$, $j\in [n]$, and $\mathfrak{v}_{(a\leftarrow u, \text{ } b\leftarrow v})$, $(u,v)\in E(G)$ and $(a,b)\in [n]\times [n]$. We construct an assignment function $f \colon V(G)\to \Sigma=[n]$ as follows. If $\mathfrak{v}_{(x,j)} \in \mathbb{U}'$, then assign $f(x)=j$. 

Any vector from $\mathbb{U}$ (or for that matter $\mathbb{V}$), has non-zero values in at most $4$ incidence dimensions, and hence its removal can affect the satisfaction in at most $4$ incidence dimensions. This implies that the vector $S(\mathbb{U}')$ has value at least $t$ in at least $2.5k-4 \delta (2.5k)$ dimensions. In all guard dimensions $S(\mathbb{U}')$ has value $1$, this implies that the vector $S(\mathbb{U}')$ has value less than $2Q$ in at most $4 \delta (2.5k)$ incidence dimensions. 

We will show that $f$ satisfies most of the constraints in $G$.  Let $e=(u,v)\in E(G)$.  Consider two vectors of form $\mathfrak{v}_{(a\leftarrow u, \text{ } b\leftarrow v)}$ and $\mathfrak{v}_{(u,f(u))}$ that belong to $U'$. If $f(u)\neq a$, then in one of the incidence dimensions, $e-u_-$ or $e-u_+$, the value of $S(U')$ is strictly less than $2Q$. Note that the only vectors with non-zero values in these dimensions are the two above vectors. For a contradiction, suppose that in both dimensions $S(U')$ has values at least $2Q$. Note that in dimension $e-u_+$, $\mathfrak{v}_{(a\leftarrow u, \text{ } b\leftarrow v)}$ has value $Q-a$ while $\mathfrak{v}_{(u,f(u))}$ has value $Q+f(u)$ and in dimension $e-u_{-}$, they have values $Q+a$ and $Q-f(u)$, respectively. Since $S(U')$ has values greater than $2Q$ in both these dimensions, it implies $(Q-a)+(Q+f(u))\geq 2Q$ and $(Q+a)+(Q-f(u))\geq 2Q$. Thus, we get $f(u)=a$, a contradiction.

We know that $S(U')$ has a value less than $2Q$ in at most $4 \delta (2.5k)$ incidence dimensions. This implies that at most $4 \delta (2.5k)$  vectors of form $\mathfrak{v}_{(a\leftarrow u, \text{ } b\leftarrow v)}$ such that either $a\neq f(u)$ or $b\neq f(v)$. This in turn implies that at most $4 \delta (2.5k)$ constraints are  not satisfied by $f$. Hence, $f$ satisfies at least $(1-\frac{20\delta}{3})\cdot \frac{3k}{2}$ constraints of $G=(V(G), E(G), \Sigma, \{C_{e}\}_{e\in E(G)})$. We choose $\delta =\frac{3\varepsilon}{20}$. This implies $(1-\frac{20\delta}{3})\cdot \frac{3k}{2}\geq (1-\varepsilon)\frac{3k}{2}$. Since $\val(G) \leq  1-\varepsilon$, this is not possible. This concludes the proof. 
\end{proof}

 \subsection{{\sc Multi-Dimensional Knapsack} to \textsc{Capacitated Vertex Cover} with multi-edges}\label{sec:reductionToUnweightedCVC}
 In this section, we give a gap-preserving reduction from {\sc Multi-Dimensional Knapsack} to \textsc{Capacitated Vertex Cover} with multi-edges, which will show that the latter problem is hard to approximate in FPT time. We start with the desired construction. 
 
 % In this section, we demonstrate a \textcolor{red}{what to call it}\emph{parameter preserving reduction}  from an instance of  {\sc Multi-Dimensional Knapsack} problem to an instance of \textsc{Muti-edge Capacitated Vertex Cover} as follows. 
%In a {\sc $d$-Dimensional Knapsack} problem, one is given an instance $(V, T, k)$, where $V = \{V_1, \cdots , V_n\}$, such that
% each $V_i \in V$ is a $d$-dimensional vector, i.e., $V_i \in Z^
% d_{Given an instance (
% \geq0}$
% ; and $T \in Z^
% d_{
% \geq0}$
% is the $d$-dimensional
% target vector. The task is to determine whether there exists a subset $U \subseteq V$ of size $k$ such that $\sum_{V_i\in U} V_i = S(U)\geq T$? For two vectors $a,b$, $a>b$ if and only if $a_i \geq b_i$ $\forall i \in[d]$ where $a=(a_1,\cdot a_d)$ and $ b=(b_1,\cdots,b_d)$. 
\medskip
%\todo{cap function should be written in some nice format}

\noindent\textbf{Construction:} Given an instance $(\mathbb{V}, \mathfrak{t}, k)$ of {\sc Multi-Dimensional Knapsack}, we construct an instance $(G,\capa,M, k+d)$ of  \textsc{Capacitated Vertex Cover} with multi-edges as follows. Reaclly that $\mathbb{V} = \{\mathfrak{v}_1, \ldots , \mathfrak{v}_n\}$, such that
each  $\mathfrak{v}_i \in \{0,1,\ldots,M\}^d$, and the target vector $\mathfrak{t} \in \{0,1,\ldots,M\}^d$. Furthermore, $d=8.5k$. 
We first define the sum of values at each coordinate. 
Let $x_i=\sum_{\mathfrak{v}\in \mathbb{V}}\alpha_i$, where $\mathfrak{v}=(\alpha_1,\cdots,\alpha_d)$. That is, $x_i$ is the sum of values at $i$-th coordinate in the input vectors. 
\begin{itemize}
\setlength{\itemsep}{-2pt}
    \item We introduce $2d$ vertices, $D=\{d_1,\cdots,d_d\}$ and $D'=\{d_1',\cdots, d_d'\}$ (two for each dimension) and a vertex $u_{\mathfrak{v}}$ in $U$ for each $\mathfrak{v}\in \mathbb{V}$, i.e, $V(G)= U\uplus D \uplus D'$.
    \item  We add edges as follows. For every $i\in[d]$, we add one edge $(d_i,d_i')$. For every vertex $u_\mathfrak{v}\in U$ where $\mathfrak{v}=(\alpha_1,\cdots,\alpha_d)$, we add $
    \alpha_i$ many edges $(u_{\mathfrak{v}},d_i)$ for every $i\in [d]$. 
    \item For each $i \in[d]$, $\capa (d_i)=x_i-t_i+1$. Here, $t_i$ is the  $i$-th coordinate of the target vector $\mathfrak{t}$.  And, the capacity of every vertex from $U$ is set to $\infty$ while the remaining vertices (from $D_i'$) have $0$ capacity.
  \item  Multiplicity function $M : V(G) \rightarrow \mathbb{N}$ assigns $1$ to each vertex. 
\end{itemize}
  
This concludes the construction. 

% In the diagram
% presented below, we outline the instance construction for a specific example.

% \begin{figure}[H]
% \includegraphics[scale=.75]{diagram-20240623.pdf}
% \caption{An illustration of construction}
% \end{figure}

\begin{theorem}\label{kanpeqv}
Unless ETH fails, there is an absolute constant
$\eta>0$, such that no algorithm that takes as input $(G, \mathsf{cap}, M, k)$  of \textsc{Capacitated Vertex Cover} with multi-edges, runs in $f(k,d)n^{\cO(1)}$ time and distinguishes between instances $(G, \mathsf{cap}, M)$ of \textsc{Capacitated Vertex Cover} with multi-edges that have a solution of size at most $k$ from instances  $(G, \mathsf{cap}, M)$having no solution of size $(1+\eta)k$. 
%The dimension of the input instance is at most $8.5k$. 
% Unless ETH fails, there is an absolute constant 
% $\eta>0$, such that no fixed parameter tractable algorithm which, takes as input $(G, \mathsf{cap}, M, k^\star)$  of \textsc{Capacitated Vertex Cover} with multi-edges, can decide whether the input is a yes instance or there is no solution of size at most $(1+\eta)k^\star$.
%This is true even when the dimension is at most $8.5k$. 
% There exists a constant $c > 1$\todo{explicit constant here?} such that assuming the ETH there is no FPT (with parameter $k$) algorithm that distinguishes between instances $(G,cap,k+d)$ of  \textsc{Multi-edge Capacitated Vertex Cover} that have a solution of size at most $k$ or have no solution of size at most $c \cdot k$.
%  \textcolor{red}{No $(1+\epsilon)$ approximate solution}

\end{theorem}
% $E(G)=\uplus_i\{(d_i,d_i') \uplus \{(d_i,x)|x\in X_i\}$ 

\begin{proof}
We first show soundness followed by completeness. 

\smallskip
\noindent 
{\bf Soundness.} Let $(\mathbb{V}, \mathfrak{t}, k)$ be a yes instance of  {\sc Multi-Dimensional Knapsack} with a solution $\mathbb{V}'\subseteq \mathbb{V}$ of size at most $k$. Let $U'=\{u_\mathfrak{v}~|~ \mathfrak{v} \in \mathbb{V}'\}$. We claim $U'\cup D$ is a solution to $(G,\capa,M,k+d)$ of  \textsc{Capacitated Vertex Cover} with multi-edges.  Here, $d=8.5k$.

Note that $S(\mathbb{V}')=\sum_{\mathfrak{v}\in \mathbb{V}'} \mathfrak{v}  \geq \mathfrak{t}$. The vertices $u_{\mathfrak{v}}$ corresponding to the vectors $\mathfrak{v}\in \mathbb{V}'$ have $\infty$ capacities. Each such vertex can cover all edges incident on it, and hence they collectively cover at least $t_i$ many edges incident on each $d_i$. Each vertex $d_i$ with its (coverage) capacity of $x_i-t_i+1$, can cover all the remaining edges incident on it from  $U'$ ($x_i-t_i$ many) as well as the edge $(d_i,d_i')$, thus making $U'\cup D$ a feasible solution for $(G, \mathsf{cap}, M, k+d)$  of \textsc{Capacitated Vertex Cover} with multi-edges. 

\smallskip
\noindent 
{\bf Completeness.} Let $(\mathbb{V}, \mathfrak{t}, k)$ be an instance of {\sc Multi-Dimensional Knapsack} such that every solution has size at least $(1+\delta)k$, here $\delta$ is an absolute constant mentioned in Theorem~\ref{cspeqv}. In this case, we will show that there exists an absolute constant
$\eta>0$, depending on $\delta$, such that the reduced instance 
$(G,\capa,M,k+d)$ of  \textsc{Capacitated Vertex Cover} with multi-edges does not have a solution of size at most $(1+\eta)(k+d)$. 

For a contradiction, assume that$(G,\capa,M,k+d)$ has a solution $V'\subseteq V(G)$ of size $(1+\eta)k$. 
% Let $(G,cap,k+d)$ instance of \textsc{Multi-edge Capacitated Vertex Cover}  have a set $V'\subseteq V(G)$ of size $(1+\epsilon)(k+d)=(1+\epsilon)20k/3$. 
Note that $D\subseteq V'$ are the only vertices capable of covering edges of the form $(d_i,d_i')$ (since the vertices in $D'$ has $0$ capacity). Thus, every vertex $d_i$ that uses one of its coverage capacity to cover $(d_i,d_i')$ is only capable of covering $x_i-t_i$ many of the edges incident on it. The rest of the $t_i$ edges incident in $d_i$ must be covered by the vertices of $V'\setminus D$, which is a subset of $U$ (as all other vertices have $0$ capacity). But this implies the existence of a set of size at most  $(1+\eta)(k+d)-d\leq k+\eta k + d \eta \leq k+ 9.5\eta k =(1+9.5\eta)k$ vertices in $V'\setminus D\subseteq U$ that covers at least $t_i$ many edges incident on $d_i$. 
This together with the fact that every vertex $u_{\mathfrak{v}}\in U$ has $\alpha_i$ many edges incident on it from $d_i$ for all $i\in [d]$, where $\mathfrak{v}=(\alpha_1,\ldots,\alpha_d)$ implies that $S(\{\mathfrak{v}~|~u_\mathfrak{v} \in V'\setminus D\})\geq \mathfrak{t}$, making $W=\{\mathfrak{v}~|~u_\mathfrak{v} \in V'\setminus D\}$ a feasible solution of {\sc Multi-Dimensional Knapsack}. If we select $\eta= \frac{\delta}{9.5}$, then we get the size of $|W|\leq (1+\delta)k$, contradicting the fact that $(\mathbb{V}, \mathfrak{t}, k)$ does not have a solution of size at most $(1+\delta)k$. This concludes the proof. 
\end{proof}

\section{Parameterized Inapproximability of {{\sc Weighted Capacitated Vertex Cover}} with multi-edges}
\label{sec:hardness-wt}
In this section, we show a tight parameterized inapproximability of  {\sc Weighted Capacitated Vertex Cover}  with multi-edges. This reduction also occurs in two steps; in the first step we reduce from {\sc $3$-Regular $2$-CSP} to {\sc Multi-Dimensional Knapsack} and in the second step we reduce {\sc Multi-Dimensional Knapsack} to {\sc Weighted Capacitated Vertex Cover}  with multi-edge.  We get a stronger hardness result for  {\sc Multi-Dimensional Knapsack} by allowing the dimension to be $d=k^{\cO(1)}$. That is, we relax the linearity condition on the number of dimensions.

% We need the constraint that $d = \cO(k)$ in the statement of Theorem~\ref{kanpeqv} because our reduction from {\sc Multi-Dimensional Knapsack} to {\sc Capacitated $d$-Hitting Set} introduces $d$ new elements that are forced into the solution to the {\sc Capacitated $d$-Hitting Set} instance. Hence a $c$-approximation for the constructed instances is allowed additive error terms that are linear in $d$.
%
% When reducing to {\sc Weighted Capacitated $d$-Hitting Set} we can give these new elements weight $0$, and thus we no longer need the requirement that $d \leq \cO(k)$. Without this additional requirement we can prove a substantially stronger hardness result for  {\sc Multi-Dimensional Knapsack}.

% \todo{say more here what is the difference..}
\subsection{Parameterized Inapproximability of {\sc Multi-Dimensional Knapsack}} 

For our reduction, we need a notion of $(\alpha,\beta,r)$-universe covering family. 

\begin{definition}
Let $U$ be a universe of size $n$ and $r\leq n$. A family $\cal F$ containing $r$ sized subsets of $U$ is called the $(\alpha,\beta,r)$-universe covering family if for every $ {\cal F}'\subseteq {\cal F}$, where $|{\cal F}'|\geq \alpha |\cal{F}|$, $|\bigcup_{A \in \mathcal{F}'}{A}|\geq (1-\beta)n$. 
\end{definition}

The next lemma shows the existence of $(\alpha,\beta,r)$-universe covering family for different parameter values of $\alpha$, $\beta$ and $r$.

\begin{lemma} \label{goodsubfamily}
For any $(\alpha,\beta)\in(0,1)$ and $r> \log_{\frac{1}{1-\beta}} (e^2/\alpha)$, there exists an $(\alpha,\beta,r)$-universe covering family of size $n/\alpha$. Here, $e$ is the base of the natural logarithm.  Furthermore, such a family can be constructed in time that depends only on $n$. 
% Given a universe $U$ of size $k$, a family $\mathcal{F}$ containing all sets of the universe of size exactly $r$, there exists a subfamily $\mathcal{F}'$ of size $ck$ such that for every subset $\mathcal{F}''\subseteq \mathcal{F}' $   of size $\delta ck$, $|\bigcup_{A_i \in \mathcal{F}''}{A_i}|\geq (1-\epsilon)k$ where $c=1/\delta$ and $r> \log_{1/1-\epsilon} (e^2/\delta)$. Moreover, such an $\mathcal{F}'$ can be found in time \textcolor{red}{fill up}.
\end{lemma}

\begin{proof}
For a family $\mathcal{Z}$ of sets over $U$, ${\sf cover}({\cal Z})$ denotes the set of elements in the union $\bigcup_{A \in \mathcal{Z}}{A}$. 
%We denote the total number of elements in a family $\mathcal{Z}$ %from a universe $U$ by $|\mathcal{Z}|_{U}$.
Let $\mathcal{F}=\{A_1,\ldots,A_{n/\alpha}\}$ be a randomly chosen family of size $n/\alpha$ from $\mathcal{F}_r$, where  $\mathcal{F}_r$ is the family containing all subsets of $U$ of size exactly $r$. If $\mathcal{F}$ is not an  $(\alpha,\beta,r)$-universe covering family, it has a subfamily $\mathcal{Z}$ of size $n$ such that  $|{\sf cover}({\cal Z})|\leq (1-\beta)n$. For this scenario, $\exists X \subseteq U$ of size $(1-\beta)n$ such that  ${\sf cover}({\cal Z})\subseteq X \subseteq U$.

\begin{align*}
\Pr{\big(\mathcal{F} \text{ is not an  $(\alpha,\beta,r)$-universe covering family}\big)} & \leq \Pr{ \big(\exists \mathcal{Z} \subseteq \mathcal{F} \text{ s.t. } |{\sf cover}({\cal Z})|\leq (1-\beta)n\big) }\\ 
          &\leq \binom{n/\alpha}{n}  \Pr(|{\sf cover}({\cal Z})|\leq (1-\beta)n)  \\&\leq \binom{n/\alpha}{n} \Pr{(\exists X \text{ s.t. } {\sf cover}({\cal Z})\subseteq X\subseteq U)}
     \\& \leq \binom{n/\alpha}{n} \cdot \binom{n}{(1-\beta)n}\cdot (1-\beta)^{nr} 
      \\&\leq  {\left(\frac{e}{\alpha}\right)^{n}} \cdot {\left(\frac{e}{1-\beta}\right)^{(1-\beta)n}} \cdot (1-\beta)^{nr}
      \\&\leq  {\left(\frac{e}{\alpha}\right)^{n}} \cdot {\left(\frac{e}{1-\beta}\right)^n} \cdot (1-\beta)^{nr} \\
      & \leq {\left(\frac{e}{1-\beta}\right)^n \cdot (1-\beta)^{nr} \cdot \left(\frac{e}{\alpha}\right)^{n}} 
     \\&\leq {(e^2/\alpha \cdot (1-\beta)^{r})}^n \hspace*{25pt} \\ 
     \\&<1
\end{align*}
The last inequality follows by substituting $r> \log_{1/1-\beta} (e^2/\alpha)$. And, it immediately implies the existence of an $(\alpha,\beta,r)$-universe covering family $\mathcal{F}$ of size $n/\alpha$. And such a family can be identified by checking if any subfamily of size $n/\alpha$ of $\mathcal{F}_r$ ($\binom{\binom{n}{r}}{n/\alpha}$ many) is an $(\alpha,\beta,r)$-universe covering family. For any constant value of $\alpha$ and $\beta$ such a family can be found in  time $\binom{\binom{n}{r}}{n/\alpha}\cdot \binom{n/\alpha}{ n} \leq n^{\mathcal{O}(n)}$.
\end{proof}

\medskip

\noindent\textbf{Construction:} 
Given an instance $G=(V(G), E(G), \Sigma, \{C_{e}\}_{e\in E(G)})$ of {\sc $3$-Regular $2$-CSP}, we construct an instance  $(\mathbb{V}, \mathfrak{t}, k^\star)$ of {\sc Multi-Dimensional Knapsack} as follows. 
We first obtain the $(\alpha,\beta,r)$-universe covering family by applying Lemma~\ref{goodsubfamily} to  the universe $V(G)$ which has size $k$. We will set $(\alpha,\beta,r)$ appropriately later. Let the family be $\mathcal{A}=\{A_1,A_2,\ldots, A_{k/\alpha}\}$ where every $A_i$ will be viewed as a set of $r$ variables $V(G)$.  We set $k^\star = \lceil \frac{k}{\alpha} \rceil$. 
We first define the number of dimensions each vector has and their significance.  The dimensions of the vectors in $\mathbb{V}$ are of the following two types.

% Given an instance $(G, C_{(u,v)\in E(G)} \subseteq [n] \times [n],k)$ of 3-regular 2-CSP problem, we construct an  instance  $(\mathbb{V}, \mathfrak{t}, k)$ of {\sc Multi-Dimensional Knapsack} as follows. Let $Q=10n$. From Lemma \ref{goodsubfamily}, we obtain a \emph{good} subfamily $\mathcal{A}=\{A_1,A_2,\ldots, A_{kc}\}$ where every $A_i$ is a set of $r$ variables from the CSP instance.
% We will claim $(G, C_{(u,v)\in E(G)} \subseteq [n] \times [n],k)$ is a yes instance if and only if there are $kc$ vectors in the {\sc $d$-Dimensional Knapsack} instance achieving the target value ($t$). 
The dimensions of the vectors in $\mathbb{V}$ are set to be of the following two types.
\begin{itemize}
\setlength{\itemsep}{-1pt}
    \item There is one \emph{guard} dimension for every $A_i\in \mathcal{A}$.
    \item For all $i,j\in[k/\alpha]$ where $i< j$, for all $u \in V(G)$ where $u\in N[A_i]\cap N[A_j]$, there are two dimensions, ${(i,j,u)}_{+}$ and ${(i,j,u)}_{-}$.
    \item Hence, the number of dimensions $d$ is upper bounded by 
   $k^\star+(k^\star)^2 \cdot k= (k^\star)^3$
\end{itemize}

Let $Q=n^{\cO(1)}$ be a {\em fixed number}. We add vectors to $\mathbb{V}$ as follows. 

\begin{itemize}
\setlength{\itemsep}{-1pt}
\item For each assignment $\gamma_i$ to $N[A_i]$ that satisfies all the edges in $N[A_i]$, we add a vector $\mathfrak{v}_{\gamma_i\leftarrow N[A_i]}$ that has 1 in the $A_i$-guard dimension, and for all $u\in N[A_i]$,

 % We add vectors to $\mathbb{V}$ as follows. For each assignment $\gamma_i$ to $N[A_i]$ that satisfies all the edges (constraints) in $N[A_i]$, we add a vector $\mathfrak{v}_{\alpha_i\leftarrow N[A_i]}$ that has $1$ in the $A_i$-guard dimension, and for all $u\in N[A_i]$,

\begin{itemize}
\item  $Q-\gamma_i(u)$ in $(i,j,u)_+$, in all those dimensions where $u\in N[A_j]$ and $i<j$,
\item  $Q+\gamma_i(u)$ in $(j,i,u)_+$, in all those dimensions where $u\in N[A_j]$ and $i>j$,
\item  $Q+\gamma_i(u)$ in $(i,j,u)_-$, in all those dimensions where $u\in N[A_j]$ and $i<j$, 
\item  $Q+\gamma_i(u)$ in $(j,i,u)_-$, in all those dimensions where $u\in N[A_j]$ and $i>j$.
\item In every other dimension, the vector has value $0$.
\end{itemize}
\item The target vector $\mathfrak{t}$ has $1$ in all guard dimensions and $2Q$ in all other dimensions.
\end{itemize}
 Note that the construction can be done in $n^{\mathcal{O}(r)}$ time. 
 %Before establishing the validity of our reduction, we introduce the notion of a %\emph{good} subfamily, demonstrate its existence, and outline the method for %identifying it in the following lemma.

%\textcolor{red}{check this theorem statement}
\begin{theorem}\label{thm:knapsack2hardness}
% Unless ETH fails, for any  constant 
% $\delta>0$, there is no fixed parameter tractable algorithm which, takes as input $(\mathbb{V}, \mathfrak{t}, k^\star)$ of {\sc Multi-Dimensional Knapsack}, can decide whether the input is a yes instance or there is no solution of size at most $(2-\delta)k^\star$.  

Unless ETH fails,  for any  constant 
$\delta >0$, such that no algorithm that takes as input $(\mathbb{V}, \mathfrak{t}, k)$  of {\sc Multi-Dimensional Knapsack}, runs in $f(k,d)n^{\cO(1)}$ time and distinguishes between instances $(\mathbb{V}, \mathfrak{t})$ of {\sc Multi-Dimensional Knapsack} that have a solution of size at most $k$ from instances $(\mathbb{V}, \mathfrak{t})$ having no solution of size $(2-\delta)k$. The dimension of the input instance is at most $k^{\cO(1)}$.   

%The dimension of input instance is at most $8.5k$. 

% For every real $c < 2$ and integer $d \geq 2$, assuming the ETH, there is no FPT (with parameter $k$) algorithm that distinguishes between instances $(\mathbb{V}, \mathfrak{t}, k)$ of {\sc Multi-Dimensional Knapsack} that have a solution of size at most $k$ and have no solution of size at least $c \cdot k$.
%$(G, C_{(u,v)\in E(G)} \subseteq [n] \times [n],k)$ is a yes instance of 3-regular-2 CSP if and only if  $(\mathbb{V},\mathfrak{t},k)$ is a yes instance of {\sc $d$-Dimensional Knapsack}.
\end{theorem}

\begin{proof}
Let $\varepsilon>0$ be an absolute constant mentioned in Proposition~\ref{prop:ETHPIH} and let $\delta>0$ be a constant. 
We set $\alpha =\delta$, $\beta =\frac{\varepsilon}{3}$, and $r=\left\lceil \log_{\frac{1}{1-\beta}} (e^2/\alpha) \right\rceil +1$ and make the reductions with these parameters. We first show soundness followed by completeness. 

\smallskip 
\noindent 
{\bf Soundness.} Let $G=(V(G), E(G), \Sigma, \{C_{e}\}_{e\in E(G)})$ be an instance of {\sc $3$-Regular $2$-CSP} such that $\val(G) = 1$. Furthermore, let $f$ be a solution assignment.  That is, $\val(G,f) = 1$. 
Let $\mathbb{U}$ be the collection of $k^\star$ vectors:
\begin{itemize}
\item one created for each $A_i\in \mathcal{A}$ with every $u\in N[A_i]$ assigned $f(u)$. That is, $$\mathbb{U}=\{\mathfrak{v}_{f \leftarrow N[A_i]}| A_i\in \mathcal{A}\}.$$

\end{itemize}

We claim $S(\mathbb{U})=\sum_{\mathfrak{v}_i\in \mathbb{U}} \mathfrak{v}_i  \geq \mathfrak{t}$. Note that in all guard dimensions, $S(\mathbb{U})$ has a value $1$ by construction. In dimensions $(i,j,u)_-$ and $(i,j,u)_+$, where $i<j$ and $u \in N[A_i] \cap N[A_j]$, the vector $\mathfrak{v}_{f\leftarrow N[A_i]}$ has values $Q-f(u)$ and $Q+f(u)$, respectively, while the vector $\mathfrak{v}_{f \leftarrow N[A_j]}$ has values $Q+f(u)$ and $Q-f(u)$, respectively. And every other vector in $\mathbb{U}$ has value 0 in these dimensions. Thus in both dimensions $S(\mathbb{U})$ has the value $2Q$. Our choices for $i,j$ and $u$ were arbitrary and hence the same arguments hold for every non-guard dimensions. This shows that $(\mathbb{V},\mathfrak{t},k^\star)$ is a yes  instance of {\sc Multi-Dimensional Knapsack}.

\medskip

\noindent 

\noindent 
{\bf Completeness.}  Let $G=(V(G), E(G), \Sigma, \{C_{e}\}_{e\in E(G)})$ be an instance of {\sc $3$-Regular $2$-CSP} such that $\val(G) \leq  1-\varepsilon$. Here, $\varepsilon>0$ is an absolute constant mentioned in Proposition~\ref{prop:ETHPIH}. In this case, we will show that for every constant
$\delta>0$, the reduced instance $(\mathbb{V},\mathfrak{t},k^\star)$ of {\sc Multidimensional Knapsack} does not have a solution of size $(2-\delta)k^\star$. 

For a contradiction, assume that $(\mathbb{V},\mathfrak{t},k^\star)$ has a set of $(2-\delta)k^\star$ vectors, say $\mathbb{U}\subseteq \mathbb{V}$ such that $S(\mathbb{U})\geq \mathfrak{t}$. To achieve the target value of $1$ in all guard dimensions, we must have a subset $\mathbb{U}'\subseteq \mathbb{U}$ containing $k^\star$ vectors. In particular, it contains vectors $\mathfrak{v}_{\gamma_i\leftarrow N[A_i]}$, $A_i\in \mathcal{A}$, and $\gamma_i$ satisfies all the edges/constraints contained in $G[N[A_i]]$.  

By the pigeonhole principle, there exist at least $\delta k^\star$ many sets in $\cal A$ for which exactly one vector $\mathfrak{v}_{\gamma_i\leftarrow N[A_i]}$ is in $\mathbb{U}$. Let these sets be denoted by ${\cal I}=\{B_1,\ldots, B_\ell\}$ and the corresponding assignments be denoted by $\gamma_i$, $i\in [\ell]$. That is, $\mathfrak{v}_{\gamma_i\leftarrow N[B_i]}$ is in $\mathbb{U}$. 
\begin{claim}
For all $i,j\in [\ell] $, if $v\in N[B_i] \cap N[B_j]$, then $\gamma_i(v)=\gamma_j(v)$. 
\end{claim}
\begin{proof}
Consider the dimensions $(i,j,v)_+$ and $(i,j,v)_-$, where $S(\mathbb{U})$ has a value of at least $2Q$. The only vectors capable of making non-zero contribution in $\mathbb{U}$ to either of these dimensions are $\mathfrak{v}_{\gamma_i\leftarrow N[A_i]}$ and $\mathfrak{v}_{\gamma_j\leftarrow N[A_j]}$. In dimensions $(i,j,u)_+$ and $(i,j,u)_-$, the values of $\mathfrak{v}_{\gamma_i\leftarrow N[A_i]}$ respectively are $Q-\gamma_i(u)$ and $Q+\gamma_j(u)$, while the values of $\mathfrak{v}_{\gamma_j\leftarrow N[A_j]}$ respectively are $Q+\gamma_j(u)$ and $Q-\alpha_j(u)$. In both dimensions $S(\mathbb{U})$ having a value at least $2Q$ implies $\gamma_i(u)=\gamma_j(u)$. This concludes the proof. 
\end{proof}

Let ${\sf extcover}({\cal I})$ denote the set of elements in the union $\bigcup_{B \in \mathcal{I}}{N[B]}$.   Now we will construct an assignment function $f \colon V(G)\to \Sigma=[n]$. If $v\in {\sf extcover}({\cal I})$  and $v\in N[B_j]$,  then $f(v)=\gamma_j(v)$.  For any other vertex, assign any arbitrary value in $[n]$. For convenience, we will assign them to $0 \notin [n]$. 

Observe that any edge $(u,v)$ where $u$ or $v$ is in ${\sf extcover}({\cal I})$, is contained in $N[B_i]$ for some $i \in[\ell]$. And $(u,v)$ is satisfied by the assignment $f$, since the vector in $\mathfrak{v}_{f\leftarrow N[B_i]} \in U$ is generated only when all edges contained in $N[B_i]$ including $(u,v)$ are satisfied by $f$.

% \begin{claim}
% Let $(u,v)$ be an edge such that $u$ or $v$ is in ${\sf extcover}({\cal I})$, then $f$ satisfies $(u,v)$. 
% \end{claim}
% \begin{proof}
% This is true because any vector in $\mathfrak{v}_{f\leftarrow N[B_i]} \in U$ is generated only when all edges/constraints in $N[B_i]$ are satisfied by $f$.
% \end{proof}

% For a vertex $v\in {\sf extcover}({\cal I})$, let ${\sf home}(v)$ be the set of indices $j\in[\ell]$ such that $v\in N[B_j]$.  An element $v\in {\sf extcover}({\cal I})$ is said to be in {\em agreement}, if for all $i,j\in {\sf home}(v)$, $\gamma_i(v)=\gamma_j(v)$.

Since, $\ell \geq \delta k^\star \geq \alpha k$, from Lemma \ref{goodsubfamily}, $|\bigcup_{i \in \mathcal{I}}{B_i}|\geq (1-\beta)k$. Observe that the number of variables not appearing in $\bigcup_{i \in \mathcal{I}}{B_i}$ is at most $\beta k$. Furthermore, the set of edges/constraints that are not satisfied is adjacent to them. This implies that the number of unsatisfied constraints is upper bounded by $3\beta k$. This implies
$ \frac{3k}{2} - 3\beta k \geq \frac{3k}{2} - \varepsilon k \geq (1-\varepsilon)\frac{3k}{2}$. Since $\val(G) \leq  1-\varepsilon$, this is not possible. This concludes the proof. 
\end{proof}

\subsection{Reducing {\sc Multi-Dimensional Knapsack} to {\sc   Weighted  Capacitated Vertex Cover}}
 In this section, we give a gap-preserving reduction from {\sc Multi-Dimensional Knapsack} to \textsc{Weighted Capacitated Vertex Cover}, which will show that the latter problem is hard to approximate in FPT time. We start with the desired construction. 
 
 % In this section, we demonstrate a \textcolor{red}{what to call it}\emph{parameter preserving reduction}  from an instance of  {\sc Multi-Dimensional Knapsack} problem to an instance of weighted \textsc{Multi-edge Capacitated Vertex Cover} as follows. 
%In a {\sc $d$-Dimensional Knapsack} problem, one is given an instance $(V, T, k)$, where $V = \{V_1, \cdots , V_n\}$, such that
% each $V_i \in V$ is a $d$-dimensional vector, i.e., $V_i \in Z^
% d_{Given an instance (
% \geq0}$
% ; and $T \in Z^
% d_{
% \geq0}$
% is the $d$-dimensional
% target vector. The task is to determine whether there exists a subset $U \subseteq V$ of size $k$ such that $\sum_{V_i\in U} V_i = S(U)\geq T$? For two vectors $a,b$, $a>b$ if and only if $a_i \geq b_i$ $\forall i \in[d]$ where $a=(a_1,\cdot a_d)$ and $ b=(b_1,\cdots,b_d)$. 
\medskip
%\todo{cap function should be written in some nice format}
%\todo{use ldots instead of cdots everywhere}

\noindent\textbf{Construction:} The construction is same as the one used in Theorem~\ref{kanpeqv} with weights being introduced. 
Given an instance $(\mathbb{V}, \mathfrak{t}, k)$ of {\sc Multi-Dimensional Knapsack}, we construct an instance $(G,\capa,M,w,k+d,W)$ of  \textsc{Weighted Capacitated Vertex Cover} with multi-edges as follows. Here, $d= k^{\cO(1)}$.  
Reacll that $\mathbb{V} = \{\mathfrak{v}_1, \ldots , \mathfrak{v}_n\}$, such that
each  $\mathfrak{v}_i \in \{0,1,\ldots,M\}^d$, and the target vector $\mathfrak{t} \in \{0,1,\ldots,M\}^d$. 
We first define the sum of values at each coordinate. 
Let $x_i=\sum_{\mathfrak{v}\in \mathbb{V}}\alpha_i$, where $\mathfrak{v}=(\alpha_1,\ldots,\alpha_d)$. That is, $x_i$ is the sum of values at $i$-th coordinate in the input vectors. 
\begin{itemize}
\setlength{\itemsep}{-2pt}
    \item We introduce $2d$ vertices, $D=\{d_1,\cdots,d_d\}$ and $D'=\{d_1',\cdots, d_d'\}$ (two for each dimension) and a vertex $u_{\mathfrak{v}}$ in $U$ for each $\mathfrak{v}\in \mathbb{V}$, i.e, $V(G)= U\uplus D \uplus D'$.
    \item  We add edges as follows. For every $i\in[d]$, we add one edge $(d_i,d_i')$. For every vertex $u_\mathfrak{v}\in U$ where $\mathfrak{v}=(\alpha_1,\cdots,\alpha_d)$, we add $
    \alpha_i$ many edges $(u_{\mathfrak{v}},d_i)$ for every $i\in [d]$. 
    \item For each $i \in[d]$, $\capa (d_i)=x_i-t_i+1$. Here, $t_i$ is the  $i$-th coordinate of the target vector $\mathfrak{t}$.  And, the capacity of every vertex from $U$ is set to $\infty$ while the remaining vertices (from $D_i'$) have $0$ capacity.
  \item  Multiplicity function $M : V(G) \rightarrow \mathbb{N}$ assigns $1$ to each vertex. 
  \item Every vertex in $U$ is assigned weight $1$, while vertices in $D$ have weight $0$. Every other vertex (in $ D'$) has weight $\infty$ (or some large number).  
  \item Set $W=k$. 
\end{itemize}
  
This concludes the construction. 

\begin{theorem}\label{kanpeqvWeighted}
Unless ETH fails, for every constant
$\eta>0$, such that no algorithm that takes as input $(G, \mathsf{cap}, M, w, k, W)$ of \textsc{Capacitated Vertex Cover} with multi-edges, runs in $f(k,d)n^{\cO(1)}$ time and distinguishes between instances $(G, \mathsf{cap}, M,w)$ of \textsc{Capacitated Vertex Cover} with multi-edges that have a solution of size at most $k$ and weight at most $W$ from instances  $(G, \mathsf{cap}, M)$ having no solution of weight at most $(2-\eta) W$.

% Unless ETH fails, there is an absolute constant 
% $\eta>0$, such that no fixed parameter tractable algorithm which, takes as input $(G, \mathsf{cap}, M, w, k, W)$  of \textsc{Capacitated Vertex Cover} with multi-edges, can decide whether the input has a solution of size at most $k$ and weight at most $W$ or has no solution of weight at most $(2-\eta) W$.

% yes instance or there is a solution of size at least $(1+\eta)k^\star$.

% that have a solution of size at most $k$ and weight at most $W$ or have no solution of weight at most $c \cdot W$.

% For every real $c < 2$ and integer $d \geq 2$, assuming the ETH, there is no FPT (with parameter $k$) algorithm that distinguishes between instances  $(G,cap,w,k+d,k_w=k)$ of weighted {\sc Multi-edge Capacitated Vertex Cover} that have a solution of size at most $k$ and weight at most $W$ or have no solution of weight at least $c \cdot W$.
 % $(\mathbb{V}, \mathfrak{t}, k)$ is a yes instance of  {\sc $d$-Dimensional Knapsack} if and only $(G,cap,k+d)$ is a yes instance  of  \textsc{Capacitated Vertex Cover}.
\end{theorem}

\begin{proof}
We first show soundness followed by completeness. 

\smallskip
\noindent 
{\bf Soundness.} Let $(\mathbb{V}, \mathfrak{t}, k)$ be a yes instance of  {\sc Multi-Dimensional Knapsack} with a solution $\mathbb{V}'\subseteq \mathbb{V}$ of size at most $k$. Let $U'=\{u_\mathfrak{v}~|~ \mathfrak{v} \in \mathbb{V}'\}$. We claim $U'\cup D$ is a solution to $(G,\capa,M,k+d,W)$ of  \textsc{Capacitated Vertex Cover} with multi-edges.  Here, $d$ could be $k^{\cO(1)}$. 

Note that $S(\mathbb{V}')=\sum_{\mathfrak{v}\in \mathbb{V}'} \mathfrak{v}  \geq \mathfrak{t}$. The vertices $u_{\mathfrak{v}}$ corresponding to the vectors $\mathfrak{v}\in \mathbb{V}'$ have $\infty$ capacities. Each such vertex can cover all edges incident on it, and hence they collectively cover at least $t_i$ many edges incident on each $d_i$. Each vertex $d_i$ with its (coverage) capacity of $x_i-t_i+1$, can cover all the remaining edges incident on it from $U'$ ($x_i-t_i$ many) and the edge $(d_i,d_i')$. Furthermore, the weight of $U'\cup D$ is at most $k$. This makes  $U'\cup D$ a feasible solution for $(G, \mathsf{cap}, M, w,k+d,k)$  of \textsc{Capacitated Vertex Cover} with multi-edges.

% ($\rightarrow$) Let $(\mathbb{V}, \mathfrak{t}, k)$ be a yes instance of  {\sc Multi-Dimensional Knapsack} with a solution $\mathbb{V}'\subseteq \mathbb{V}$. We claim $U'\cup D$ is a solution to $(G,cap,w,k+d,k)$ of weighted \textsc{Multi-edge Capacitated Vertex Cover} where $U'=\{u_\mathfrak{v}| \mathfrak{v} \in \mathbb{V}'\}$. Note that $\sum_{\mathfrak{v}\in \mathbb{V}'} \mathfrak{v} = S(\mathbb{V}')\geq \mathfrak{t}$. The vertices $u_{\mathfrak{v}}$s corresponding to the vectors $v\in \mathbb{V}'$ have $\infty$ capacities. Each such vertex can cover all edges incident on it, and hence they collectively cover at least $t_i$ many edges incident on each $d_i$. Each vertex $d_i$ with its (coverage) capacity of $x_i-t_i+1$, can cover all the remaining edges incident on it from  $U'$ ($x_i-t_i$ many) as well as the edge $(d_i,d_i')$, thus making $U'\cup D$ (a set of size $k+d$ and weight $k$) a feasible solution for $(G,cap,w,k+d,k)$ of \textsc{Multi-edge Capacitated Vertex Cover} problem.  

\smallskip
\noindent 
{\bf Completeness.} Let $(\mathbb{V}, \mathfrak{t}, k)$ be an instance of {\sc Multi-Dimensional Knapsack} such that every solution has size at least $(2-\delta)k$, here $\delta$ is an absolute constant mentioned in Theorem~\ref{thm:knapsack2hardness}. In this case, we will show that there exists an absolute constant
$\eta>0$, depending on $\delta$, such that the reduced instance 
$(G,\capa,M,w,k+d,W)$ of  \textsc{Capacitated Vertex Cover} with multi-edges does not have a solution of  at most $(2-\eta) W$.

For a contradiction, assume that $(G,\capa,M,w,k+d,W)$ have an inclusion-wise minimal set $V'\subseteq V(G)$ of weight $(2-\epsilon)k$.  Note that $D\subseteq V'$ are the only vertices capable of covering edges of the form $(d_i,d_i')$ (since the vertices in $D'$ has $0$ capacity). Thus, every vertex $d_i$ that uses one of its coverage capacity to cover $(d_i,d_i')$ is only capable of covering $x_i-t_i$ many of the edges incident on it. The rest of the $t_i$ edges incident in $d_i$ must be covered by the vertices of $V'\setminus D$, which is a subset of $U$ (as all other vertices have $0$ capacity). But this implies the existence of a set of weight at most $(2-\eta) W=(2-\eta) k$ in $V'\setminus D\subseteq U$ that covers at least $t_i$ many edges incident on $d_i$.

This together with the fact that every vertex $u_{\mathfrak{v}}\in U$ has $\alpha_i$ many edges incident on it from $d_i$ for all $i\in [d]$, where $\mathfrak{v}=(\alpha_1,\ldots,\alpha_d)$ implies that $S(\{\mathfrak{v}~|~u_\mathfrak{v} \in V'\setminus D\})\geq \mathfrak{t}$, making $Z=\{\mathfrak{v}~|~u_\mathfrak{v} \in V'\setminus D\}$ a feasible solution of {\sc Multi-Dimensional Knapsack} of {\em size} $(2-\eta) k$.  If we select $\eta= \delta$, then we get the size of $|Z|\leq (2-\eta) k$, contradicting the fact that $(\mathbb{V}, \mathfrak{t}, k)$ does not have a solution of size at most $(2-\delta)k$. This concludes the proof. 
% ($\leftarrow$) Let $(G,cap,w,k+d,k)$ of  {\sc Multi-edge Capacitated Vertex Cover} have an inclusion-wise minimal set $V'\subseteq V(G)$ of weight $(2-\epsilon)k$. Note that $V'$ being a inclusion-wise minimal does not contain any vertex with 0 capacity. And $D\subseteq V'$, as the only vertex capable of covering a $(d_i,d_i')$ edge is $d_i$ (since $d_i'$ has $0$ coverage capacity). Any $d_i$ spending one of its coverage capacity to cover  $(d_i,d_i')$ is only capable of covering $x_i-t_i$ many of edges incident on it. 
% The rest of $t_i$ edges incident on $d_i$ must be covered by the vertices in $V'\setminus D$ which is a subset of $U$ (as all other vertices have $0$ coverage capacities). The rest of $t_i$ edges incident on $d_i$ must be covered by the vertices in $V'\setminus D$ which is a subset of $U$ (as all other vertices have $0$ capacities). But this implies the existence of a set of weight $(2-\epsilon)k$  in $V'\setminus D\subseteq U$ that covers at least $t_i$ many edges incident on $d_i$. This together with the fact that every  vertex $u_{\mathfrak{v}}\in U$ has $v_i$ many edges incident on it from $d_i$, where $\mathfrak{v}=(v_1,\ldots,v_d)$ implies that $S(\{\mathfrak{v}|u_\mathfrak{v} \in V'\setminus D\})\geq \mathfrak{t}$, making $\{\mathfrak{v}|u_\mathfrak{v} \in V'\setminus D\}$ a feasible solution of  $(2-\epsilon)k$ vectors.
\end{proof}

\section{Conclusion}
\label{sec:conclusion}
We gave a parameterized approximation algorithm for the \textsc{Capacitated $d$-Hitting Set} problem. The algorithm outputs a single solution which simultaneously achieves an approximation ratio of $4/3$ for the size of the solution, and $2+\epsilon$ for the weight.

Our approximation algorithm can also be made to work for the related \textsc{Partial Capacitated $d$-Hitting Set} problem. Here input additionally comes with an integer $t$ and the task is to cover at least $t$ sets using the minimum number of elements. Indeed the reduction of Cheung et al.~\cite{DBLP:conf/soda/CheungGW14} that reduces \textsc{Partial Capacitated $d$-Hitting Set} to \textsc{Capacitated $(d+1)$-Hitting Set} combined with our algorithm does the job. Note that, as opposed to {\em polynomial time} approximation algorithms for these problems our approximation ratio does not depend on $d$. Thus, from the perspective of FPT-approximation \textsc{Partial Capacitated $d$-Hitting Set} appears no harder than  \textsc{Capacitated $d$-Hitting Set}.

We also showed that, assuming the ETH, for {\em size} no FPT-approximaition algorithm can achieve an approximation ratio arbitrarily close to $1$, and for {\em weight} no FPT-approximaition algorithm can achieve an approximation ratio arbitrarily below $2$. Hence our algorithm achieves the best possible approximation ratio one can hope for in terms of weight, and is not too far off in terms of size. On the way to showing the hardness result for \textsc{Capacitated $d$-Hitting Set} we also obtain a tight factor $2-\epsilon$ inapproximability result for FPT-approximating {\sc Multi-Dimensional Knapsack} which may be of independent interest. 

It is an intriguing open problem whether it is possible to close the gap and identify the smallest $c$ such that unweighted \textsc{Capacitated $d$-Hitting Set} admits an  
FPT-approximaition algorithm with factor $c$, and prove hardness (under ETH or another plausible assumption) of $(c-\epsilon)$-approximation. This would likely also lead to improved understanding of FPT-approximation and FPT-inapproximability of fundamental parameterized problems such as {\sc 3-Regular 2-CSP}.

\bibliographystyle{alpha}
\bibliography{main.bib}
\end{document}